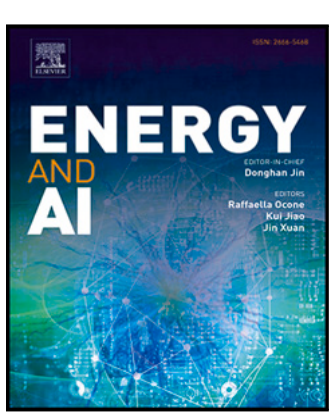

# Degradation-aligned self-supervised learning for state of health estimation of lithium-ion batteries under label sparsity

Jiaqi Yao *, Julia Kowal

*Department of Electrical Energy Storage Technology (EET), Technische Universität Berlin, Einsteinufer 11, 10587, Berlin, Germany*

## HIGHLIGHTS

- A ranking-based SSL framework is proposed for SOH estimation under label sparsity.
- A CNN-GRU model is developed for SOH estimation with local feature fusion.
- Insights are presented on the influences of label distribution of degradation data.

## GRAPHICAL ABSTRACT

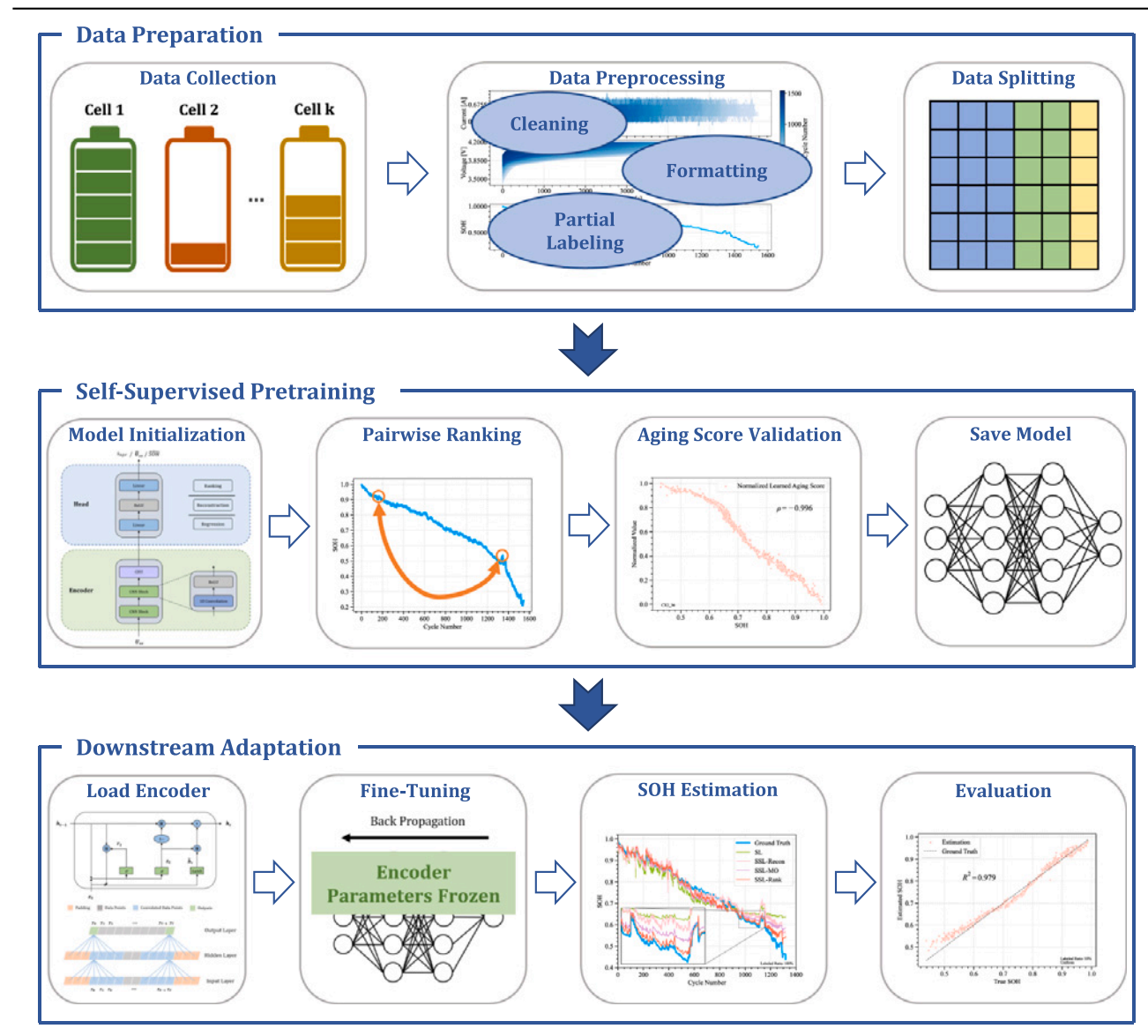

## ARTICLE INFO

## ABSTRACT

An accurate estimation of the state of health (SOH) underpins safe and optimized use of the battery system. Although compelling, data-driven SOH estimation models typically require large amounts of high-quality labeled cycling data, while in practice such labels are often sparse in both quantity and coverage. Therefore, in this work, we propose a degradation-aligned self-supervised learning (SSL) framework based on a convolutional neural network-gated recurrent unit (CNN-GRU) model, which learns aging-consistent representations from unlabeled data through a cycle-order ranking objective as the pretext task for pretraining, thereby enabling robust SOH estimation after fine-tuning on sparsely labeled data. Test results showcase that the proposed ranking-based SSL approach proves to endow the pretrained model with degradation awareness from unlabeled data, and after fine-tuning the model can carry out accurate, robust SOH estimation, even when only an extremely limited amount of 1% of unevenly distributed labeled training data is available, where the MAE of 1.718% and RMSE of 2.329% can be achieved on the test cell. In addition, in-depth analyses are presented regarding the influences of label distribution and cross-cell robustness. We believe this work could shed new light on label-efficient SOH estimation of lithium-ion batteries, addressing a practical need in battery management.

* Corresponding author.
*E-mail address:* jiaqi.yao@tu-berlin.de (J. Yao).

## 1. Introduction

The growing urgency of global climate change has significantly accelerated the transition to clean, sustainable energy systems [1–4]. Renewable energy sources such as wind [5] and solar power [6,7] have been rapidly developed and deployed worldwide. However, their inherent intermittency and volatility pose significant challenges to the stability and reliability as an energy supply, thereby creating a strong demand for dependable, efficient energy storage systems [8]. Benefiting from many advantages, including high energy and power densities [9,10], low self-discharge [11], and longevity [12], lithium-ion batteries have been widely utilized across numerous fields, from portable consumer electronics [13,14] to electric transportation [15,16]. Regardless of the specific application, an accurate determination of the state of health (SOH) is indispensable for lithium-ion batteries and is a core functionality of battery management systems (BMSs). At some time $t$, the SOH of a battery can be defined either based on the capacity or the internal resistance [17,18]:

$$SOH_C(t) = \frac{C_{actual}(t)}{C_{rated}} \tag{1}$$

$$SOH_R(t) = \frac{R_{actual}(t) - R_{EOL}}{R_{rated} - R_{EOL}} \tag{2}$$

where $C_{actual}(t)$ and $C_{rated}$ respectively denote the actual battery capacity at time $t$ and the rated capacity, while $R_{actual}(t)$, $R_{EOL}$, and $R_{rated}$ respectively denote the actual battery internal resistance at time $t$, the predefined end-of-life (EOL) internal resistance, and the rated internal resistance. Usually, the capacity-based definition of $SOH_C$ is utilized without further clarification, as in this work. SOH is a defined metric for quantifying battery degradation. The determination of SOH plays a vital role in battery systems [19–21], as it not only provides essential information for the assessment of remaining useful life (RUL) and the early scheduling of necessary maintenance, but also underpins a safe and optimized use of the battery by supporting the prevention of unexpected failures and improving the overall lifespan utilization of the battery systems.

However, SOH is closely related to the complex underlying electrochemical processes and can hardly be directly measured; thus, it has to be estimated from the measurable information during operation [22]. In general, SOH estimation approaches can be categorized into three types [23,24]: direct measurement approaches, model-based approaches, and data-driven approaches. Direct measurement approaches estimate the SOH based on the knowledge acquired from measurements, such as deriving the actual usable capacity by applying coulomb counting on a standardized discharge capacity test [25]. Although straightforward and easy to implement, such approaches can hardly be applied in real applications, as the fully controlled operating conditions and complete discharge cycles required are generally unavailable in practice. Model-based SOH estimation approaches can be further categorized into two groups, namely degradation modeling approaches and parameter identification approaches [24]. Degradation modeling approaches exploit various models that describe battery degradation, such as empirical models [26,27] and electrochemical models [28,29], to infer the current usable capacity, while parameter identification approaches cast the problem of SOH estimation as a parameter identification problem utilizing battery models like equivalent circuit models (ECMs) [30] and electrochemical models [31], and nonlinear state observers like extended Kalman filters (EKFs) [32], unscented Kalman filters (UKFs) [33], and particle filters [34]. Despite the fact that model-based approaches are generally considered accurate and robust, in-depth domain knowledge is required for the modeling of the electrical, aging, and even thermal behavior of the battery, and specifically designed characterization tests are compulsory to parameterize these models.

In recent years, data-driven solutions have come into the spotlight for various applications. Especially in the field of deep learning, prosperous progresses are being made. In the context of battery SOH estimation, data-driven approaches aim to capture the underlying degradation patterns directly from operational data without requiring prior knowledge of battery electrochemical dynamics. However, most existing works choose to manually extract input features for data-driven models due to the belief that handcrafted health indicators can more explicitly characterize battery degradation and thus reduce the difficulty of model learning with increased interpretability [35]. In Ref. [36], the authors proposed a lightweight local health indicator extraction approach for multi-stage fast charging protocols, where the segmented charging data of certain state of charge (SOC) windows was used as the input features, which were then fed into a hybrid deep learning model for the mapping of SOH estimates. The proposed approach achieved precise SOH estimation, with mean absolute errors (MAEs) and root mean square errors (RMSEs) below 1%. In Ref. [37], the authors proposed an improved gated recurrent unit (GRU) network combined with the whale optimization algorithm for SOH estimation, where the time duration in the voltage ranges from 3.55 V to 3.75 V and from 3.80 V to 4.15 V during constant current (CC) charging was extracted as the two input features. Test results showed that the proposed method is able to achieve an average error of less than 1% and presents good generalization capability. In Ref. [38], the authors conducted a comprehensive characterization of the aging behavior patterns of lithium-ion batteries, where five aging patterns were identified, covering the rate of voltage change during discharging and CC charging, the duration of CC charging, the rate of current change during constant voltage (CV) charging, and the rate of temperature change during CV charging. These aging patterns were then used as health indicators for the input of deep neural networks (DNNs). Test results showed that the proposed health indicators consistently improved estimation performance across different DNN-based SOH estimators. In Ref. [39], incremental capacity analysis (ICA) was utilized for feature extraction, where the incremental capacity differences of five voltage intervals were selected as the input feature candidates of a multi-layer perceptron (MLP) for SOH and RUL estimation. Test results showed that the proposed approach could achieve relative error rates below 3% for SOH estimation. In Ref. [40], the authors focused on real-life electric vehicle (EV) driving scenarios and proposed a snapshot-based approach with a long short-term memory (LSTM) network for flexible SOH estimation, where the voltage and current sequences of partial charging segments with a fixed window size were used as the model input. In addition, the authors fused the snapshot-based and conventional history-based approaches to develop a noise-robust approach. Test results showed that the proposed approaches were able to achieve an average error of less than 2.46% over all presented experiments. In Ref. [41], on the other hand, the authors directly took the full charging curves of voltage, current, and temperature as the input features of the proposed GRU-convolutional neural network (CNN) for SOH estimation. Test results on the National Aeronautics and Space Administration (NASA) dataset and the Oxford battery degradation dataset showcased that the proposed approach was able to achieve accurate SOH estimation with the maximum estimation error of less than 4.3%, demonstrating the fact that deep learning models are capable of automatically extracting useful features from the measured cycling data and mapping them into accurate SOH estimates.

However, data-driven SOH estimation approaches rely on large amounts of high-quality labeled battery cycling data, where the SOH labels are typically obtained through standardized checkup tests that cover the entire lifespan of the batteries under controlled conditions. Such calibration tests are time-consuming and expensive for laboratories, and can seldom be conducted in real-world deployment. Furthermore, late-life SOH labels are inherently more difficult to collect, as reaching the deep-degradation regime requires extensive cycling over a long period of time with considerable experimental costs, resulting in a pronounced imbalance in label distribution across the battery lifespan. As a result, in practical application scenarios, battery aging data often face the problem of label sparsity, both in quantity and coverage, while large volumes of unlabeled operational cycling data are available. Self-supervised learning (SSL), a branch of machine learning, is specifically aimed at such challenges. SSL methods enforce the model to learn meaningful data representations without relying on manual labels by exploiting underlying data structures or automatically constructed pretext tasks derived from the unlabeled data itself [42–44]. In general, SSL techniques can be categorized into four types [44]: generative approaches, which learn representations by reconstructing the input data or predicting the missing parts of it [45,46]; contrastive approaches, which learn representations by forcing similar samples to be closer and dissimilar samples to be more distant in the latent space [47,48]; contrastive generative approaches, which combine both generative and contrastive objectives and exploit the advantages of both [49,50]; context-based approaches, which learn representations by leveraging contextual information of the samples, such as temporal order and spatial structures [51,52]. In fact, SSL techniques have been sparsely applied in some previous works in the field of battery SOH estimation to address the problem of label sparsity. In Ref. [53], the authors proposed an SSL framework utilizing a reconstruction-based generative approach with an auto-encoder–decoder, aiming to address the problems of limited labeled data and underutilized measurements during degradation. In their work, an auto-encoder–decoder was trained to reconstruct the partial capacity–voltage curve as the pretext task. After the SSL pretraining, the trained encoder was transferred to the downstream network for SOH estimation, which was then fine-tuned using the sparse labeled data. Test results showed that the proposed framework was able to achieve a robust, accurate SOH estimation using a very limited amount of labeled data. In Ref. [54], the authors proposed a self-supervised framework incorporating weak labels to reduce the demand for large amounts of annotated battery aging data for deep learning-based SOH estimation approaches, where the raw data was first processed into three-dimensional feature maps with enriched information. Afterwards, the authors exploited the generative pretext tasks of masked image reconstruction and charging capacity estimation for pretraining. Similar to the aforementioned work, a small amount of labeled data was then used for fine-tuning on the downstream task of SOH estimation. Test results showed that the proposed model demonstrated strong generalization even with only a few labeled data points. On the other hand, in Ref. [55], the authors proposed a multi-level contrastive SSL approach with dynamic embedding, aiming to leverage the multi-level physical temporal dependence of the data to solve the problem of label scarcity. Frequency domain information was extracted through the discrete Fourier transform, after which the embedding was obtained using their proposed multi-scale dynamic embedding method. Contrastive learning combined with mask reconstruction was utilized for SSL pretraining. Finally, the pretrained model was fine-tuned using limited labeled data for SOH estimation. A summary of the existing works is shown in Table 1.

As a matter of fact, current SSL approaches in the field of SOH estimation tend to focus on reconstruction-based pretext tasks, where the model is trained to recover the input or its masked parts, as such strategies help the model capture general underlying patterns of the data. However, the learned representations are not necessarily aligned with the battery degradation process that is most relevant to the task of SOH estimation. In contrast, the relative aging order between cycles provides a much more direct and task-relevant self-supervised signal, since battery aging is inherently an ordered process. Therefore, in this work, we propose an intuitive degradation-aligned SSL framework that learns aging-consistent representations from unlabeled CC charging curves through a cycle-order ranking objective as the pretext task for pretraining, thereby enabling robust SOH estimation after fine-tuning on only a limited amount of labeled data for practical scenarios where data labels are sparse in both quantity and coverage. In addition, we develop a CNN-GRU model that extracts local patterns from charging curves via convolutional layers and then integrates these features sequentially through recurrent units for accurate SOH estimation as well as self-supervised pretraining. To the best of our knowledge, this work is the first to exploit the intrinsic information behind the cycle order on battery aging for self-supervised SOH estimation. The main contribution of this work is as follows:

1. A ranking-based SSL framework is proposed to learn degradation-consistent representations from unlabeled CC charging curves via cycle-order ranking as pretraining for robust SOH estimation in scenarios where only sparsely labeled data are available for fine-tuning.
2. A CNN-GRU model is developed to extract and sequentially integrate the local patterns from charging curves for accurate SOH estimation as well as self-supervised pretraining.
3. In-depth analyses are presented regarding the influences of label distribution of battery degradation data.

The rest of this paper is structured as follows: Section 2 introduces the proposed SSL framework for SOH estimation of lithium-ion batteries under label sparsity, with a detailed explanation of the developed model, the ranking-based SSL algorithm, and the overall workflow. Section 3 describes the experimental setups, including data preparation, settings of the presented models and experiments, and the utilized evaluation metrics. In Section 4, comprehensive results of a variety of experiments are presented, together with in-depth analyses. The key takeaways of this work are summarized in Section 5.

## 2. Methodology

### 2.1. CNN-GRU model for self-supervised pretraining and SOH estimation

#### 2.1.1. Convolutional neural networks

CNNs [56] are a class of deep learning models that apply trainable convolution kernels with local receptive fields and shared weight parameters to extract local patterns from structured data. Fig. 1 shows the working scheme of an example two-layer 1D CNN with a kernel size of $k = 7$. The dilation factor $d$ and the stride $s$ are assumed to be 1. In order to preserve the input length after each convolution operation, the same padding technique is adopted in the network. Specifically, when the kernel size $k$ is an odd number, the padding size on both sides of the 1D sequence is given by:

$$p = \frac{k-1}{2} \tag{3}$$

With the same padding technique, the output sequence will have the same length as the input sequence without being shortened. In this case, the output vector at time step $t$, namely $\boldsymbol{y}_t$, is dependent on the inputs $\boldsymbol{x}_{t-p}, \boldsymbol{x}_{t-p+1}, \dots, \boldsymbol{x}_t, \dots, \boldsymbol{x}_{t+p-1}, \boldsymbol{x}_{t+p}$, where the inputs with indices outside the original index range of the input sequence are the appended paddings. More generally, at some time step $t$, the 1D convolution operation under same padding $G(t)$ can be formulated as:

$$G(t) = (\boldsymbol{X} * \boldsymbol{F})(t) = \sum_{i=0}^{k-1} \boldsymbol{F}(i) \cdot \boldsymbol{x}_{t+i-p} \tag{4}$$

where $\boldsymbol{X}$ is the input sequence and $\boldsymbol{F}$ denotes the convolution kernel with kernel size $k$. CNNs effectively capture local patterns in input data with relatively few parameters, which is why they are widely used across tasks such as image recognition, object detection, and signal processing. In our context of battery SOH estimation, the cycling curves of batteries can also be viewed as structured one-dimensional signals, where adjacent measurements often exhibit strong local correlations. Since battery degradation gradually alters the shape and patterns of the cycling measurement trajectories, CNNs are a natural choice for the extraction of local patterns correlated with aging from such data, which is why they are utilized as part of the network backbone in this work for SOH estimation as well as the self-supervised pretraining.

**Table 1**
Comparison of existing and our works on SSL-based SOH estimation.

| Reference | Methodology | Objective |
|---|---|---|
| Ref. [53] | Reconstruction-based generative approach with auto-encoder–decoder | To address the problems of limited labeled data and underutilized aging measurements |
| Ref. [54] | SSL framework incorporating weak labels with a five-layer Transformer encoder for feature extraction | To reduce the demand for large amounts of annotated battery aging data |
| Ref. [55] | Multi-level contrastive SSL approach with dynamic embedding | To leverage the multi-level physical temporal dependence of the data to solve label scarcity |
| Our work | Degradation-aligned SSL framework utilizing a cycle-order ranking objective for pretraining | To enable robust SOH estimation after fine-tuning on sparsely labeled data in real-world applications |

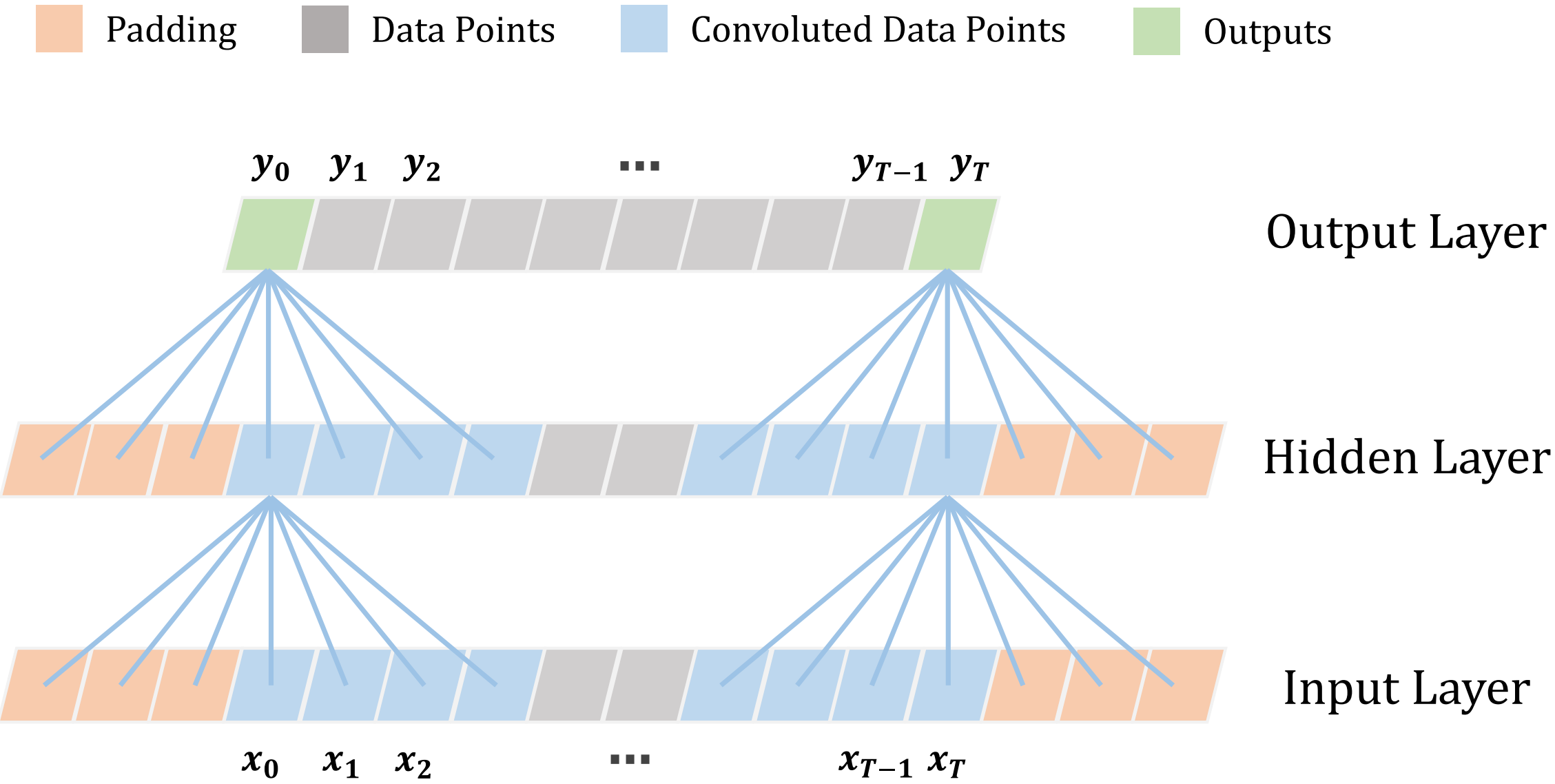

**Fig. 1.** Working scheme of an example two-layer 1D CNN.

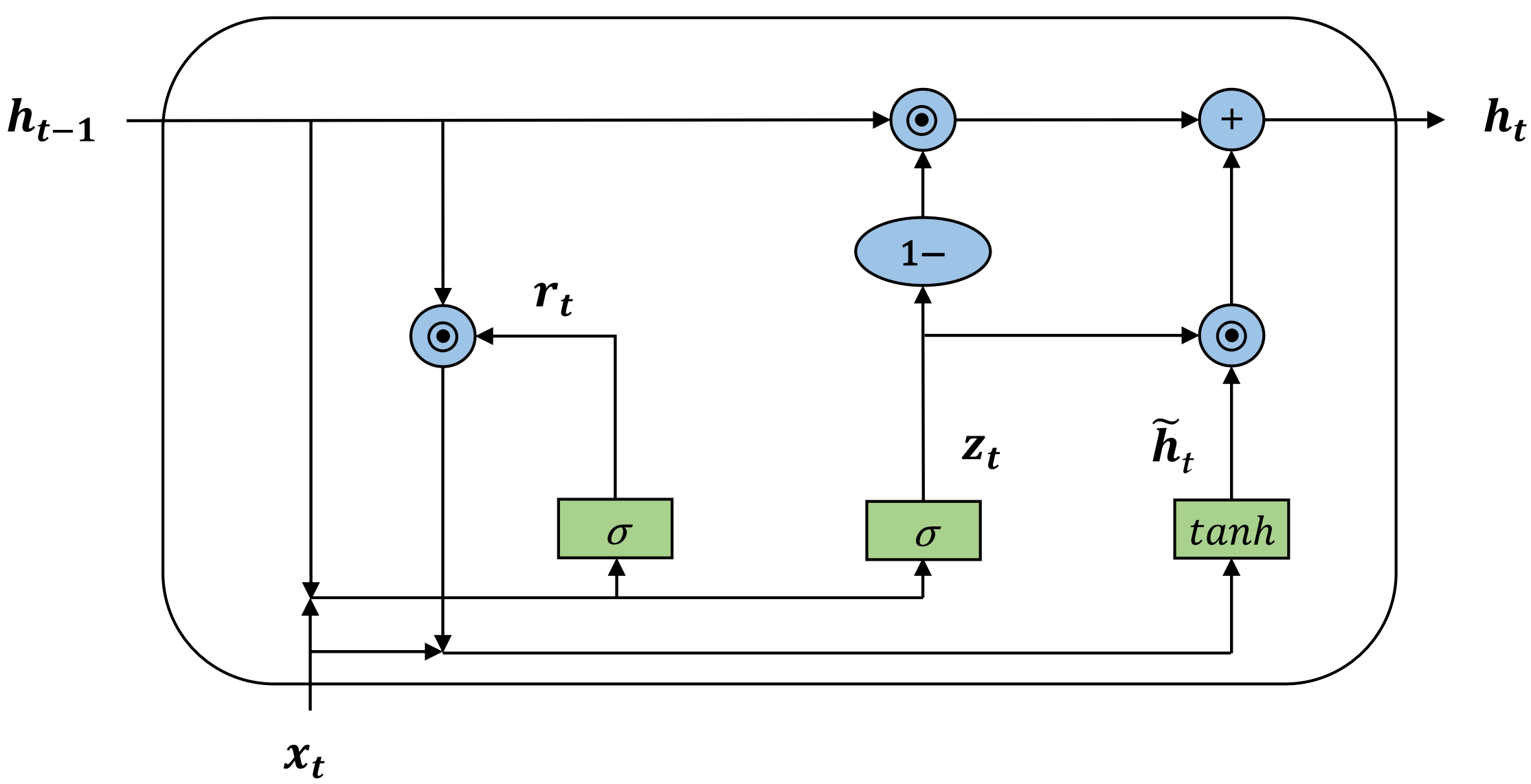

**Fig. 2.** Internal architecture of GRU.

#### *2.1.2. Gated recurrent units*

GRUs [57] are an advanced variant of recurrent neural networks (RNNs) that are designed for sequential dependency modeling while mitigating the problem of vanishing gradient in vanilla RNNs. With the introduction of gating mechanisms, GRUs can regulate the information flow actively and update the hidden state selectively based on the importance of historical and current information. In addition, the lean internal architecture of GRUs facilitates high computational efficiency and lower dependencies on the amount of required training data, especially when compared with the other frequently applied variant of RNN with gating mechanisms, namely LSTM [58], making them a perfect choice for our case of SOH estimation under label sparsity. The internal architecture of the GRU is demonstrated in Fig. 2. At time step $t$, given the input vector $\boldsymbol{x}_t$, the update gate $\boldsymbol{z}_t$ and reset gate $\boldsymbol{r}_t$ can be calculated as:

$$\boldsymbol{z}_t = \sigma(\boldsymbol{W}_{xz}\boldsymbol{x}_t + \boldsymbol{W}_{hz}\boldsymbol{h}_{t-1} + \boldsymbol{b}_z) \tag{5}$$

$$\boldsymbol{r}_t = \sigma(\boldsymbol{W}_{xr}\boldsymbol{x}_t + \boldsymbol{W}_{hr}\boldsymbol{h}_{t-1} + \boldsymbol{b}_r) \tag{6}$$

where $\sigma(\cdot)$ denotes the sigmoid activation function, $\boldsymbol{h}_{t-1}$ denotes the hidden state of the previous time step, and $\boldsymbol{W}$ and $\boldsymbol{b}$ denote the respective weight matrix and bias vector. The candidate hidden state $\tilde{\boldsymbol{h}}_t$ can thereby be computed by:

$$\tilde{\boldsymbol{h}}_t = tanh(\boldsymbol{W}_{xh}\boldsymbol{x}_t + \boldsymbol{W}_{hh}(\boldsymbol{r}_t \odot \boldsymbol{h}_{t-1}) + \boldsymbol{b}_h) \tag{7}$$

where $\odot$ denotes the Hadamard product, namely element-wise multiplication. In the end, the final output of the hidden state $\boldsymbol{h}_t$ is updated as:

$$\boldsymbol{h}_t = (1 - \boldsymbol{z}_t) \odot \boldsymbol{h}_{t-1} + \boldsymbol{z}_t \odot \tilde{\boldsymbol{h}}_t \tag{8}$$

The reset gate controls the forgetting and retention of the historical information for the calculation of the candidate hidden state. The candidate hidden state combines the current input and the selected historical information. The new hidden state is the final output at the current time step from the fusion of the previous hidden state and candidate hidden state controlled by the update gate. In this work, the GRU is utilized to sequentially integrate the local patterns extracted by the previous CNN from CC charging curves, thereby comprehensively capturing the underlying patterns correlated with battery degradation for self-supervised representation learning and downstream SOH estimation.

#### 2.1.3. CNN-GRU model

Fig. 3 shows the architecture of the proposed CNN-GRU model. The model consists of two parts, namely the encoder and head. The encoder aims to learn the underlying patterns from the input charging curves and is composed of two stacked CNN blocks and one GRU layer, where each CNN block further consists of one 1D convolutional layer and one rectified linear unit (ReLU) as the activation function. The CNN is responsible for extracting local patterns from the input charging curves, and the GRU is used to sequentially integrate the extracted features so as to capture their temporal dependencies. Based on the learned latent representation of the encoder, the head is further used to accomplish the target task, either self-supervised pretraining or SOH regression, which is composed of two fully connected (FC) layers and ReLU in between as the activation function. The same head architecture is used for both self-supervised pretraining and SOH estimation, but of course with different loss functions and sizes for the outputs. For the proposed ranking-based self-supervised pretraining, the head is used to generate an aging score $s_{age}$ based on the learned representations. For reconstruction-based self-supervised pretraining, the head is used to reconstruct the respective input charging curve $\hat{\boldsymbol{U}}_{cc}$ based on the embedded vector generated by the encoder. For the downstream SOH estimation task, the head performs regression for $\hat{SOH}$. We use the CC segment of charging as input, since, in contrast to the highly dynamic discharge process with substantial fluctuations in the operating conditions, the charging process generally follows a clearer protocol and is acquired under more stable conditions for EVs and electronics [36,40]. As a result, it is more feasible to charge data to build high-quality, large-scale datasets that are continuously accumulated across the entire usage history, with strong comparability across different cycles. In addition, during the CC charging stage, the curve shape often exhibits evident pattern shifts with battery aging, making it a strong indicator of battery health. Since previous works have showcased that deep learning models are able to extract the aging-related features from the charging curve automatically [41], we use the voltage measurement of the CC charging segment $\boldsymbol{U}_{cc}$ directly as the input of the model. It is worth mentioning that the inclusion of other health indicators, such as the temperature response or manually extracted statistical features, could improve robustness in more diverse operating conditions.

### 2.2. Cycle-order ranking for degradation-aligned SSL

**Algorithm 1** Computation of intra-cell cycle-order ranking loss for a sampled mini-batch.

**Require:** Mini-batch $\{(c_i, n_i, s_i)\}_{i=1}^{B}$, where $c_i$ is the cell ID, $n_i$ is the cycle number, and $s_i$ is the predicted aging score of sample $i$; minimum cycle gap $d_{\min}$

1: $\mathcal{G} \leftarrow \text{GroupByCell}(\{1,\dots,B\}; \{c_i\}_{i=1}^{B})$ ▷ group sample indices by cell ID
2: $\mathcal{S} \leftarrow \emptyset$ ▷ store group-wise ranking losses
3: **for all** $G_m \in \mathcal{G}$ **do**
4: **if** $|G_m| < 2$ **then**
5: **continue** ▷ at least two samples are needed
6: **end if**
7: $\tilde{G}_m \leftarrow \text{Permute}(G_m)$ ▷ generate a shuffled index order for subsequent pairing
8: $\mathcal{P}_m \leftarrow \{(i,j) \mid i \in G_m,\ j \in \tilde{G}_m,\ n_i \neq n_j,\ |n_i - n_j| \geq d_{\min}\}$ ▷ construct sample pairs with ambiguous pairs removed
9: **if** $|\mathcal{P}_m| = 0$ **then**
10: **continue** ▷ skip if no valid pair remains
11: **end if**
12: $y_{ij} \leftarrow \begin{cases} +1, & n_i > n_j \\ -1, & n_i < n_j \end{cases}, \quad \forall (i,j) \in \mathcal{P}_m$ ▷ later cycles should receive larger aging scores
13: $\mathcal{L}_m \leftarrow \frac{1}{|\mathcal{P}_m|} \sum_{(i,j)\in\mathcal{P}_m} \ln\big(1 + \exp(-y_{ij}(s_i - s_j))\big)$ ▷ group-wise logistic ranking loss
14: $\mathcal{S} \leftarrow \mathcal{S} \cup \{\mathcal{L}_m\}$ ▷ collect valid group losses
15: **end for**
16: **if** $|\mathcal{S}| = 0$ **then**
17: **return None** ▷ no valid ranking supervision in this mini-batch
18: **else**
19: $\mathcal{L}_{\text{rank}} \leftarrow \frac{1}{|\mathcal{S}|} \sum_{\mathcal{L}_m \in \mathcal{S}} \mathcal{L}_m$ ▷ average over valid cell groups
20: **return** $\mathcal{L}_{\text{rank}}$
21: **end if**

Battery aging is an inherently progressive process, in which later cycles generally correspond to deeper degradation states than earlier ones. Such an ordinal relationship provides a natural, intuitive source of self-supervision even in the absence of explicit SOH labels. Motivated by this observation, we introduce an intra-cell cycle-order ranking objective for self-supervised pretraining to encourage the model to learn degradation-aligned representations from unlabeled charging data. Specifically, for each sampled mini-batch during training, the samples are first grouped according to their cell IDs, and ranking is only performed within each cell group. This design avoids unreliable comparisons across different cells, whose degradation trajectories may differ due to inter-cell variability. Within each cell group, random sample pairs are constructed for efficient optimization. In order to improve the reliability of the ranking signal, pairs with identical cycle numbers or with cycle-number differences smaller than a predefined threshold $d_{\min}$ are excluded, since their degradation order is either undefined or too weak to provide clear supervision due to possible capacity recovery phenomena. For a valid pair of samples $(i, j)$ from the same cell, the ranking label is defined as:

$$y_{ij} = \begin{cases} +1, & \text{if } n_i > n_j \\ -1, & \text{if } n_i < n_j \end{cases} \tag{9}$$

where $n_i$ and $n_j$ denote the corresponding cycle numbers. Here, $y_{ij} = +1$ indicates that sample $i$ comes from a later cycle and is expected to have a larger aging score than sample $j$, as the pairwise logistic ranking loss is designed to be:

$$\ell_{ij} = \ln\left(1 + \exp\left(-y_{ij}(s_i - s_j)\right)\right) \tag{10}$$

where $s_i$ and $s_j$ are the aging scores predicted by the model under training. It is referred to as a logistic ranking loss because it is derived from the negative log-likelihood of a logistic model, in which the probability of a correct pairwise order $p_{ij}$ is modeled as a sigmoid function of the margin $y_{ij}(s_i - s_j)$, namely:

$$p_{ij} = \frac{1}{1 + \exp\left(-y_{ij}(s_i - s_j)\right)} \tag{11}$$

where the margin indicates both whether the predicted order is correct and how confidently the model makes this prediction. Minimizing the negative log-likelihood of this probability $p_{ij}$ yields the loss $\ell_{ij}$, which encourages the score difference $s_i - s_j$ to be consistent with the relative cycle order. For the $m$th valid cell group, with $\mathcal{P}_m$ denoting the set of valid sample pairs constructed within that group, the group-wise ranking loss is then computed as:

$$\mathcal{L}_m = \frac{1}{|\mathcal{P}_m|} \sum_{(i,j)\in\mathcal{P}_m} \ell_{ij} \tag{12}$$

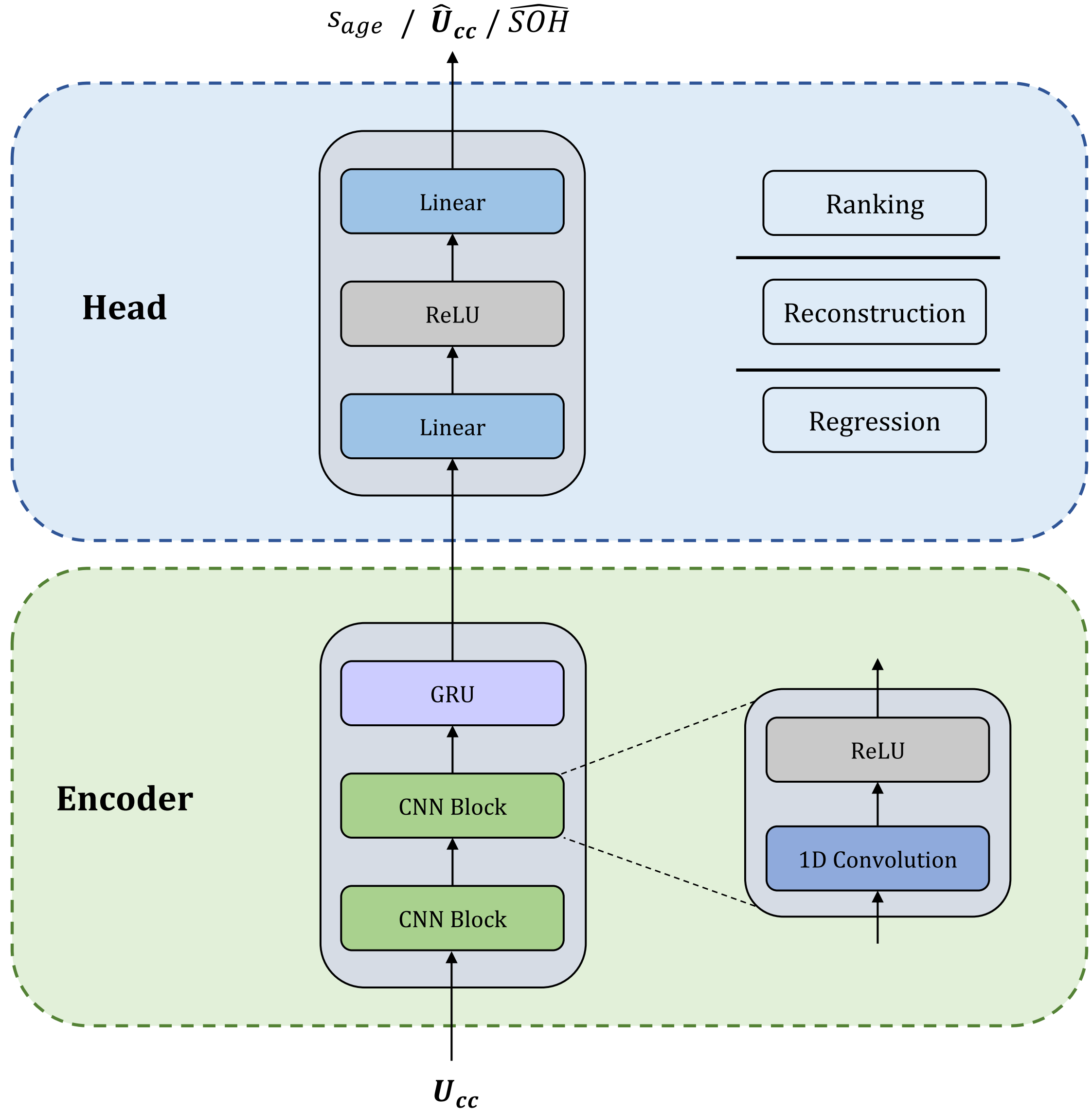

**Fig. 3.** The proposed CNN-GRU model for self-supervised pretraining and SOH estimation.

Finally, the overall ranking loss for the mini-batch is obtained by averaging over all valid cell groups:

$$\mathcal{L}_{\text{rank}} = \frac{1}{M} \sum_{m=1}^{M} \mathcal{L}_m \tag{13}$$

where $M$ is the number of valid cell groups in the mini-batch. Algorithm 1 summarizes the intra-cell cycle-order ranking objective for a sampled mini-batch.

During ranking-based self-supervised pretraining, the model, consisting of the encoder and the head, is expected to predict a dimensionless aging score from the CC charging curve input. By explicitly enforcing that the predicted aging scores follow the relative degradation order across cycles, the proposed ranking objective encourages the encoder to capture the underlying patterns correlated to the degradation process from unlabeled charging curves. In this way, the learned representations become better aligned with the evolution of battery degradation, which is closely related to the downstream SOH estimation task, thereby benefiting cases with label sparsity in both quantity and coverage.

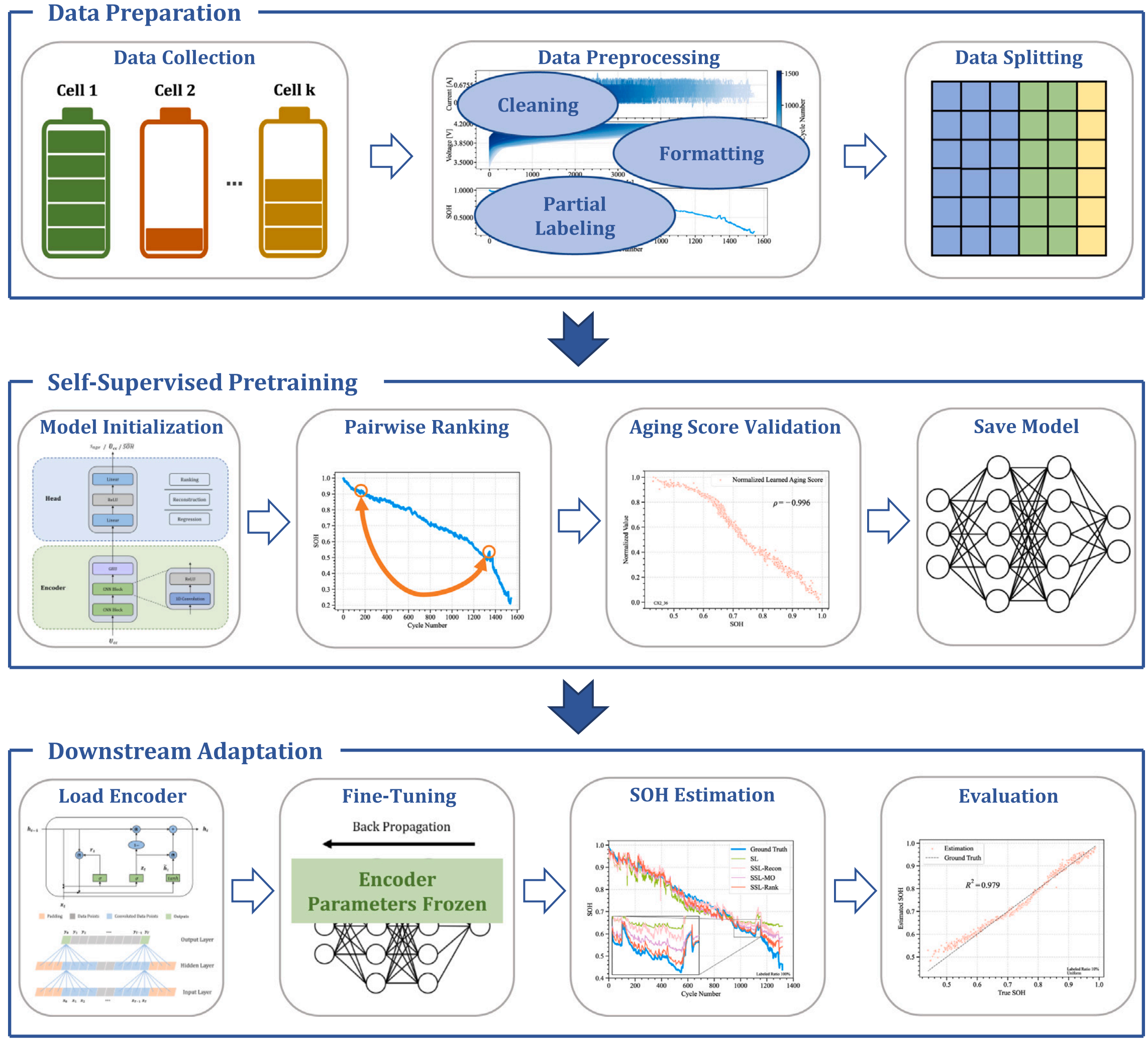

**Fig. 4.** Workflow of the proposed SSL framework for SOH estimation under label sparsity.

### 2.3. Degradation-aligned SSL framework

Fig. 4 shows the overall workflow of the proposed SSL framework for SOH estimation under label sparsity. First, as data preparation, the collected battery cycling aging data undergo a series of preprocessing procedures, including cleaning of dirty samples and outliers, formatting for a unified structure, and labeling the data partially based on the experiment settings to simulate practical scenarios where the battery aging data are under label sparsity, both in quantity and coverage. The preprocessed data are then split into training, validation, and test sets for parameter optimization, hyperparameter fine-tuning, and objective evaluation, respectively. The unlabeled data will be used for self-supervised pretraining, while the partially labeled subset will be used for fine-tuning the pretrained model during downstream adaptation. During self-supervised pretraining, the CNN-GRU model introduced in Section 2.1, composed of an encoder and a head, is first initialized. Afterwards, the model is trained on the pretext task with unlabeled charging data using the ranking-based SSL algorithm proposed in Section 2.2. The outcome of the pretraining can be validated by inspecting the correlation between the pretrained model's predicted aging score and the ground-truth SOH values, if available. The pretrained model will be saved and transferred to downstream adaptation on the task of SOH estimation, where only the pretrained encoder will be loaded, and a new head will be instantiated for SOH regression. Subsequently, the labeled subset of data will be used for fine-tuning the model, during which only the new head is updated, with the parameters of the pretrained encoder frozen. The fine-tuned model is ready for accurate, robust SOH estimation on new cells despite label sparsity in the training data.

To clarify the distinction between the various types of data and their uses, we provide a more detailed explanation of the data pipeline. The data pipeline consists of four distinct parts. First, unlabeled cycling data are used for self-supervised pretraining, where only cycle-order information is used to construct ranking pairs and no SOH labels are required. Second, a sparse subset of labeled cycles from the training cells is used for supervised fine-tuning. Third, the validation cell with available SOH labels is used only for hyperparameter selection and model selection. Finally, the test cell is strictly held out and its SOH labels are used only for the final evaluation.

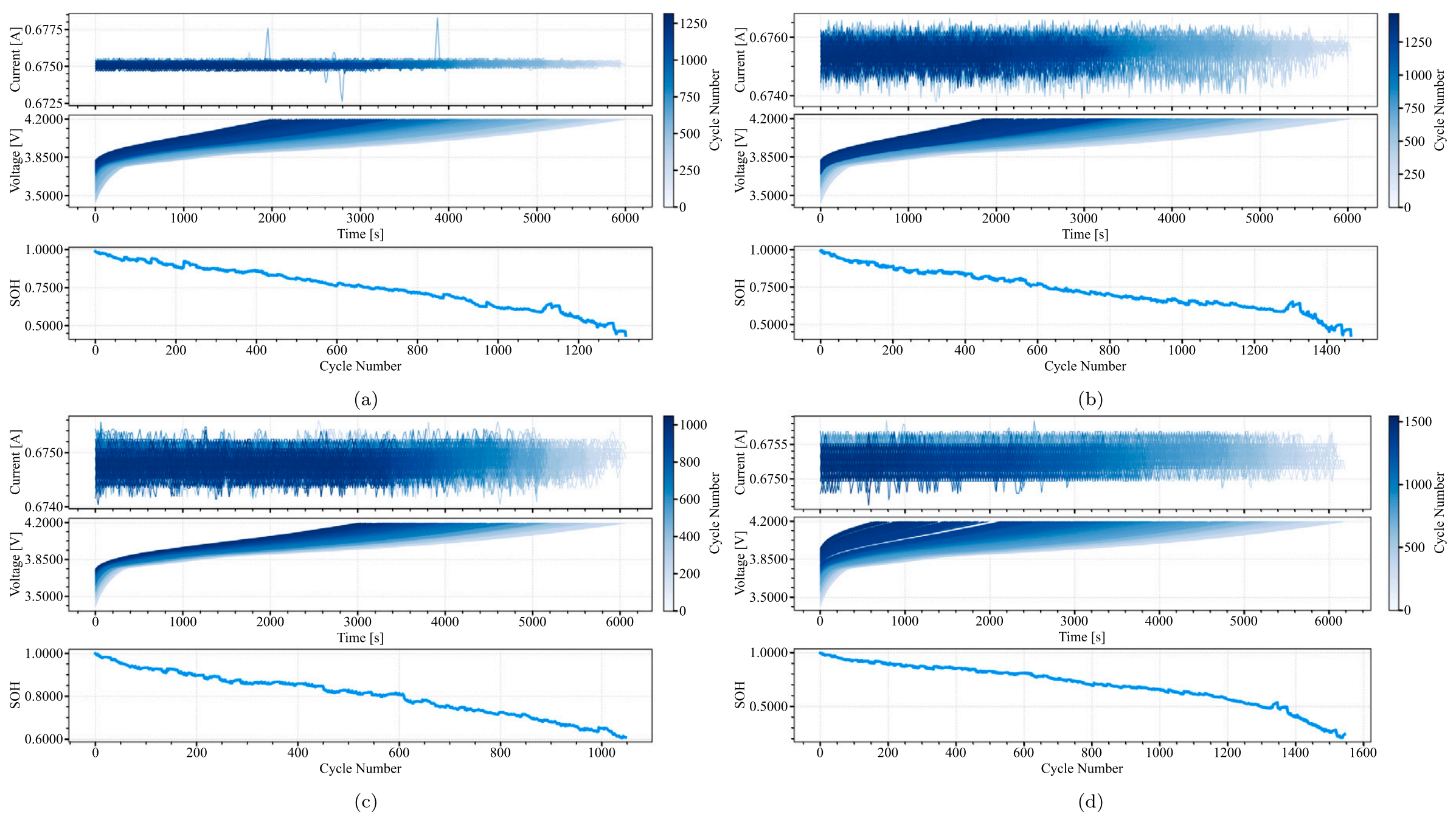

**Fig. 5.** Aging profiles of the cells, including the current and voltage measurement during CC charging and the evolution of SOH. (a) Cell CX2-34. (b) Cell CX2-36. (c) Cell CX2-37. (d) Cell CX2-38.

**Table 2**
Specifications of the utilized CX2 battery.

| Parameter | Data |
|---|---|
| Nominal capacity | 1.35 Ah |
| Cell chemistry | LCO |
| Cell format | Prismatic |
| End-of-charge voltage | 4.20 V |
| End-of-discharge voltage | 2.70 V |

## 3. Experimental setup

### 3.1. Data preparation

In this work, the experiments and analyses are based on the public battery aging dataset, which is widely used in the research community, from the Center for Advanced Life Cycle Engineering (CALCE) at the University of Maryland [59]. Specifically, the cells CX2-34, CX2-36, CX2-37, and CX2-38 are utilized in this work, of which the specifications are shown in Table 2. These prismatic cells have lithium cobalt oxide (LCO) cathodes and a rated capacity of 1.35 Ah. The four cells underwent cyclic aging from the beginning of life until different depths of degradation. For charging, they went through a CC-CV protocol, where the cells were charged under the constant current rate of 0.5 C until the end-of-charge voltage of 4.2 V was reached, after which CV charging carried on until the charging current dropped below 0.05 A. For discharging, the cells went through a CC protocol with a current rate of 1 C [18] until the end-of-discharge voltage of 2.7 V was reached.

BatteryML [60], an open-source platform for machine learning-based battery degradation diagnostics, is used for preliminary preprocessing of the raw cycling data, including the initial integration and ordering of the measurement files, compilation of basic information, and rough cleaning of dirty entries. Afterwards, the voltage measurements of the CC charging fragment are extracted for each cycling curve and resampled to a fixed length of 300 points to facilitate later training and to enforce the network to learn the underlying patterns from the partial charging curve rather than simply using the sequence length as the indicator. The corresponding SOH labels are calculated based on Eq. (1) using the full discharge capacity. Consequently, more thorough data cleaning is performed, removing outliers based on anomalies in current and voltage readings, as well as on the rolling median absolute deviation (MAD) of the lengths of CC charging segments at the cell level. More specifically, samples with CC charging segments of less than 5 data points are first cleaned, and then samples whose last current reading during CC charging is smaller than 0.6 A or the first voltage reading during CC charging is larger than 4 V are removed as outliers. Afterwards, rolling MAD of a moderate window size 31 and threshold 3 is applied onto the SOH degradation curves for further outlier removal. The window size of 31 samples is used to estimate the local trend. The residuals between the original signal and this rolling median are then computed, and their MAD is scaled by 1.4826 to estimate the noise level. Samples whose robust z-score exceeded 3 are treated as outliers. The preprocessed aging profiles of the cells are shown in Fig. 5. As can be observed, the overall shape and pattern of the voltage measurement during CC charging demonstrate an obvious shift through the aging process: the initial voltage reading becomes higher due to impedance rise, and the duration of the CC fraction becomes shorter due to the combination of loss of lithium inventory (LLI), loss of active material (LAM), and impedance rise, which intuitively underpins the reliability of using voltage measurement during CC charging as the input feature for SOH estimation. The capacity degradation trends of the cells appear similar, where the four cells are aged to different depths of degradation. Based on this observation, CX2-37 and CX2-38 are selected for training because together they cover relatively early and late degradation stages, enabling the model to learn from a wider range of aging conditions. CX2-36 and CX2-34, which fall between these two extremes in terms of degradation depth, are used for validation and testing, respectively. In this way, the validation and test cells remain unseen during training, while their degradation states are still enclosed by the training distribution rather than lying outside it. This split is therefore intended to enable a more stable and meaningful evaluation of cross-cell generalization. However, in order to test the model's robustness across different testing scenarios, we have also conducted further experiments using a variety of data splits in Section 4.4.

### 3.2. Experiment settings

Z-score normalization is utilized on the training set to accelerate convergence in this work using the following formula:

$$z = \frac{x - \mu}{\sigma} \tag{14}$$

where $z$ is the normalized value, $x$ is the raw value, $\mu$ is the mean, and $\sigma$ is the standard deviation. The utilized CNN-GRU models all have the same architectural hyperparameters regardless of whether they are for self-supervised pretraining or SOH estimation. For the encoder part, the two-layer CNN uses a slightly larger kernel size of 7 to capture local patterns across more adjacent points and increases the number of channels from 1 to 32 and then to 64. Replicate values are used for the same padding on both sides of the input sequence. After the CNN, a one-layer GRU with an input size of 64 and an output size of 128 is used to fuse the extracted local patterns. For the head part, the fully connected layers have different numbers of neurons for different tasks. For ranking-based self-supervised pretraining and for SOH regression, the two fully connected layers decrease the size of the latent vector gradually from 128 to 64, and then to 1, yielding a scalar of aging score or SOH estimate. For the reconstruction pretext task, which is used as a comparison in the experiments, the two fully connected layers gradually increase the size of the latent vector from 128 to 256, and then to 300. The batch size of 256 is used in this work. Automatic hyperparameter fine-tuning of the learning rate is conducted for all models, including both self-supervised pretraining and fine-tuning, using the hyperparameter optimization framework Optuna [61], utilizing the tree-structured Parzen estimator (TPE) algorithm, based on their performance on the validation set for a fair comparison.

The ranking-based self-supervised pretraining uses a minimum cycle gap of $d_{min} = 50$ to mitigate disturbances to the supervision signal caused by the capacity recovery effect during the aging test. Besides the proposed ranking-based SSL (SSL-Rank) approach, we apply two other SSL approaches for pretraining as comparison, including the most frequently used reconstruction-based SSL (SSL-Recon) approach and a multi-objective SSL (SSL-MO) approach. As the pretext task, SSL-Recon aims to reconstruct the voltage curves during CC charging through the encoder and the head, thereby encouraging the model to learn informative representations that preserve structural characteristics of the input charging signals. And as the name indicates, SSL-MO combines reconstruction and ranking as its pretext task in order to test the possibility that these two self-supervised objectives can complement each other and thereby yield more informative representations for downstream SOH estimation. The SSL models are first pretrained using the respective SSL strategy on the unlabeled data, after which a new head will be instantiated on top of the pretrained encoder for adaptation using the limited labeled data. During the adaptation, the pretrained encoder will be frozen, meaning only the freshly instantiated head will be updated. Certainly, at the same time, we train ordinary supervised learning (SL) models with the same network architecture using the limited labeled data for SOH estimation as a baseline. In addition, the conventional machine learning approach, ridge regression, is used as a simple baseline for extra analyses. Specifically, a ridge regression model with the cycle number as input (Ridge-C) is trained to serve as a direct indicator of the correlation between SOH and the cycle number, and a ridge regression model with the CC charging voltage as input (Ridge-V) is trained as a leaner model that serves as a fairer baseline in such data-sparse scenarios. It is worth mentioning that the SL model refers specifically to the neural network with the same architecture trained under SL in this work, despite the fact that ridge regression generally also belongs to SL. For the automatic fine-tuning using Optuna, all models follow the same tuning budget: for all the deep learning models, namely the SSL pretraining models, the fine-tuning models after pretraining, and the SL models trained from scratch, grid searches of the optimized learning rate are carried out in the range from $1e{-}5$ to $1e{-}2$ for 50 trials; for the ridge regression models, grid searches of the optimized regularization penalty factor $\alpha$ are carried out in the range from $1e-6$ to $1e3$ for 50 trials. All searches use the minimization of the validation error on the validation set as the objective, where for the SSL pretraining, it is the respective unlabeled SSL loss, and for SOH estimation, it is the labeled estimation MAE.

### 3.3. Evaluation metrics

The regression performance of the SOH estimation task is evaluated using the following four evaluation metrics in this work, namely MAE, RMSE, maximum absolute error (MAX), and the coefficient of determination $R^2$:

$$MAE(\boldsymbol{y}, \hat{\boldsymbol{y}}) = \frac{1}{n}\sum_{i=1}^{n}|y_i - \hat{y}_i| \tag{15}$$

$$RMSE(\boldsymbol{y}, \hat{\boldsymbol{y}}) = \sqrt{\frac{1}{n}\sum_{i=1}^{n}(y_i - \hat{y}_i)^2} \tag{16}$$

$$R^2(\boldsymbol{y}, \hat{\boldsymbol{y}}) = 1 - \frac{\sum_{i=1}^{n}(y_i - \hat{y}_i)^2}{\sum_{i=1}^{n}(y_i - \bar{y})^2} \tag{17}$$

$$MAX(\hat{\boldsymbol{y}}) = \max\{|y_i - \hat{y}_i|\}_{i=1}^{n} \tag{18}$$

where $\boldsymbol{y}$ and $\hat{\boldsymbol{y}}$ denote the labels and the estimates, respectively. $n$ is the number of samples, while $y_i$ and $\hat{y}_i$ denote the label and the estimate of the $i$th sample, respectively. $\bar{y}$ is the mean of the labels $\boldsymbol{y}$. The four evaluation metrics for regression have different emphases. MAE is an intuitive measure for the average magnitude of estimation errors, while RMSE gives greater weight to large errors due to the squaring operation. $R^2$ is used to evaluate how well the estimations capture the overall variation trend of the labels during fitting, and MAX reflects the worst-case estimation deviation across all samples. In this work, SOH is shown in its normalized fractional form between 0 and 1, which is the direct output/target representation used in the model. In contrast, the evaluation errors are reported in percentage points to improve readability and to make the magnitude of the errors easier to interpret. For example, an MAE of 0.005 in fractional SOH corresponds to 0.5% in the reported metrics.

In addition, Spearman's rank correlation coefficient $\rho$ is employed to evaluate the monotonic relationship between the predicted aging scores and the ground-truth SOH values after the self-supervised pretraining, thereby assessing whether the learned representations can consistently reflect the real degradation process. It is defined as:

$$\rho = \frac{\mathrm{cov}(\boldsymbol{R}_s, \boldsymbol{R}_y)}{\sigma_{R_s}\sigma_{R_y}} \tag{19}$$

where $\boldsymbol{R}_s$ and $R_y$ denote the ranks of the predicted aging scores and the ground-truth SOH labels, respectively. cov denotes covariance, and $\sigma_{R_s}$ and $\sigma_{R_y}$ denote the standard deviations of the corresponding rank variables. $\rho$ is defined in the range from $-1$ to 1, where $\rho = 1$ indicates a perfect positive monotonic relationship, and $\rho = -1$ indicates a perfect negative monotonic relationship.

## 4. Results and discussion

In this section, we present the results and analyses of a variety of experiments conducted to evaluate the pretrained model via degradation alignment, compare different SSL and SL approaches on sparsely labeled data, and examine the influence of label distribution.

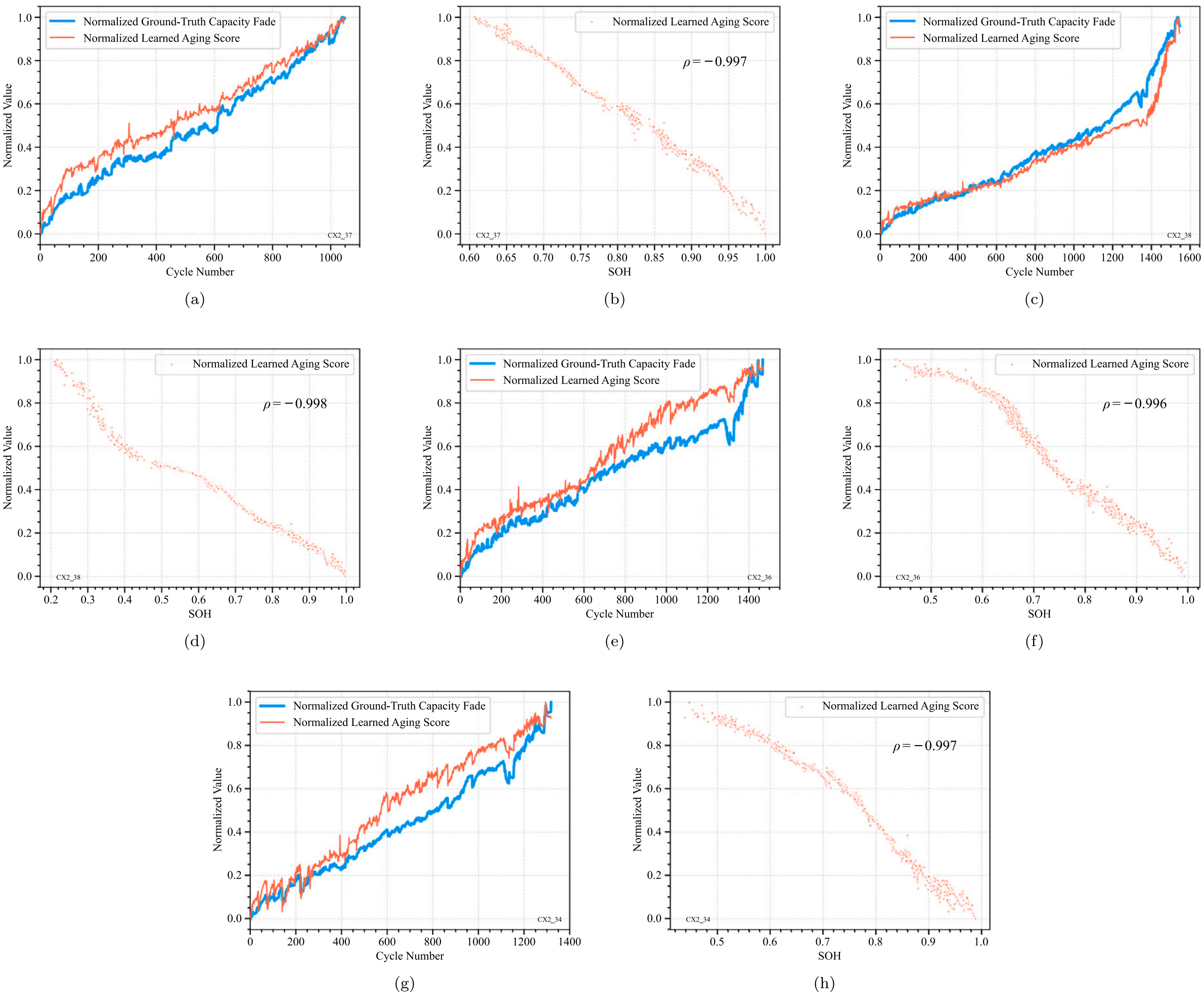

**Fig. 6.** Correlation between the aging score predicted by the pretrained model under ranking-based SSL and the respective ground-truth degradation process. (a) Cell CX2-37, normalized aging score and capacity fade. (b) Cell CX2-37, normalized aging score and SOH. (c) Cell CX2-38, normalized aging score and capacity fade. (d) Cell CX2-38, normalized aging score and SOH. (e) Cell CX2-36, normalized aging score and capacity fade. (f) Cell CX2-36, normalized aging score and SOH. (g) Cell CX2-34, normalized aging score and capacity fade. (h) Cell CX2-34, normalized aging score and SOH.

### *4.1. Evaluation of the pretrained model via degradation alignment*

In this experiment, we aim to evaluate the pretrained model using the proposed ranking-based SSL approach before downstream fine-tuning. To this end, the pretrained model is used to generate an aging score for each input charging curve, and the resulting scores are compared with the corresponding ground-truth degradation process in order to verify their alignment. Since the proposed ranking objective is designed to encode the relative degradation order of battery cycles into the learned representations and then map it into an unscaled aging score prediction, effective pretraining is expected to produce aging scores that vary monotonically with SOH. Therefore, both the overall relationship between the predicted aging scores and the degradation process, and their Spearman's rank correlation coefficient are analyzed to assess whether the pretrained model has successfully captured degradation-aligned information from unlabeled charging data.

Fig. 6 shows the correlation between the aging score predicted by the pretrained model under the proposed ranking-based SSL and the ground-truth degradation processes of the four cells, namely CX2-37, CX2-38, CX2-36, and CX2-34, across the training, validation, and testing sets. Since the learned aging score of the pretrained model is an unscaled value without a unit, we first normalize it into the range from 0 to 1 for better visualization, as well as the ground-truth capacity fade, using the following formula of min–max normalization:

$$x' = \frac{x - x_{min}}{x_{max} - x_{min}} \tag{20}$$

where $x'$ is the normalized value, $x$ is the original value, and $x_{max}$ and $x_{min}$ denote the maximum and minimum of the variable, respectively. In the left column of Fig. 6, namely Figs. 6(a), 6(c), 6(e), and 6(g), we compare the aforementioned normalized aging scores with the ground-truth capacity fade over the entire lifespan of each cell. As can be observed, despite the minor local fluctuations, the aging scores predicted by the pretrained model generally demonstrate a clear monotonic overall trend that is well aligned with the capacity degradation trajectory of each cell. In particular, as the battery ages and the capacity gradually fades, the predicted aging scores consistently evolve in the corresponding direction, indicating that the pretrained model has successfully captured degradation-relevant information from unlabeled charging data. This is particularly evident for cells CX2-37 and CX2-38, namely the two training cells, as the ranking objective was directly imposed on them during the training process. Especially for the training cell CX2-38, the aging scores predicted by the pretrained model can even reconstruct the different degradation paces of the ground-truth degradation process, showing clearly different slopes before and after the 1350th cycle. For the validation cell CX2-36, from around the 600th cycle, the predicted aging score begins to show slight divergence from the ground-truth capacity fade, although the monotonicity and alignment are not compromised, and the dip indicating capacity recovery at around the 1300th cycle is well captured by the pretrained model. Similarly, on the test cell CX2-34, the predicted aging score generally aligns quite well with the ground-truth capacity fade, despite local fluctuations at the beginning of life (BOL) and a slight divergence in the middle of life. The capacity recovery phenomenon at around the 1150th cycle is well captured by the pretrained model based on the unlabeled CC charging data as well.

In the right column of Fig. 6, namely Figs. 6(b), 6(d), 6(f), and 6(h), the correlation between the normalized aging score and the ground-truth SOH is showcased for each cell, with Spearman's rank correlation coefficient calculated and marked in each case. The normalized aging scores generally exhibit a clear monotonic decreasing trend as the ground-truth SOH increases, or equivalently, a monotonic increasing trend as the battery degrades. This indicates that the predicted aging scores are strongly consistent with the underlying aging progression of each cell. Although a few outliers can still be observed locally, the overall relationship remains highly correlated. The calculated Spearman's rank correlation coefficients $\rho$, namely −0.997 for cell CX2-37, −0.998 for cell CX2-38, −0.996 for cell CX2-36, and −0.997 for cell CX2-34, are all extremely close to −1, indicating an almost perfect monotonic inverse relationship between the predicted aging score and the ground-truth SOH for all cells, which further supports the fact that our proposed ranking-based SSL approach is able to enforce the pretrained model to extract degradation-aligned information from unlabeled charging data, such that the learned aging scores consistently reflect the underlying battery aging progression across the entire lifespan. Still, it is worth mentioning that testing the learned aging score on cells with more pronounced regeneration events or highly irregular degradation trajectories would be a useful way to further verify whether it reflects true degradation rather than simply cycle number.

Under the current experimental setting, we have conducted a quantitative analysis of the number of ranking pairs discarded due to the violation of the minimum cycle gap $d_{min}$. Specifically, pairs with identical cycle numbers or with cycle-number differences smaller than 50 were removed from the ranking loss. Under this setting, in each epoch, 243 out of 2593 candidate pairs were discarded, corresponding to 9.371% of all candidate pairs. In general, a smaller $d_{min}$ retains more ranking pairs but may introduce ambiguous or noisy supervision from close-cycle pairs, especially in the presence of local capacity recovery. Conversely, a larger $d_{min}$ increases the reliability of the ranking labels by enforcing a clearer degradation separation, but it also reduces the number of available training pairs. Considering the fact that exhaustive tuning of this hyperparameter would take an extremely long time, as all the pretraining and fine-tuning processes are affected, $d_{min} = 50$ is selected as a conservative trade-off between pair reliability and pair availability.

### 4.2. Model performance comparison under label sparsity

In this experiment, we aim to compare the performance of the models pretrained using different SSL strategies after fine-tuning, as well as under complete SL with no pretraining on the unlabeled data. More specifically, we consider three different SSL approaches: SSL-Rank, our proposed SSL approach using the cycle-order objective as the pretext task; SSL-Recon, the most frequently used SSL approach using reconstruction of the input CC charging curves as the pretext task; SSL-MO, which combines reconstruction and ranking as its pretext task in order to test the possibility that these two self-supervised objectives can complement each other and thereby yield more informative representations for downstream SOH estimation. All models are first pretrained on the unlabeled CC charging data using their respective SSL strategies, where they have achieved the expected good performance, and then fine-tuned using sparsely labeled CC charging data. In addition, we train a pure SL model from scratch without pretraining on the same sparsely labeled CC charging data using the same network architecture, and two ridge regression models, namely Ridge-C and Ridge-V, with the cycle number and the CC charging voltage as input, respectively. Here, we consider a sparse-label setting in which only data above the 80% SOH threshold are labeled, and the proportion of labeled samples varies from 1% to 100%, where 100% of labeled data corresponds to approximately 1200 training samples. This setup is consistent with practical scenarios, since 80% SOH is commonly adopted as the EOL criterion for lithium-ion batteries, and thus data from this earlier degradation region are more likely to be available than labeled samples from deep-degradation stages, whose acquisition requires much longer cycling time and higher experimental cost. The labeled data points for fine-tuning or complete SL are sampled uniformly from all data points within the defined SOH range for each training cell.

Fig. 7 shows the SOH estimation results on the test cell CX2-34 of different models after fine-tuning or complete training on different ratios of labeled data. At first glance, we can see that all models pretrained using SSL strategies are able to carry out satisfactory SOH estimation under different ratios of labeled data from 1% to 100% on the entire lifespan of the test cell, while the pure SL model without pretraining shows great fluctuations in its estimation under ratios of labeled data above 50% and fails the task of SOH estimation under ratios of labeled data below 40%, because the number of samples available for SL from scratch becomes too small for the model to capture the underlying battery degradation dynamics. Compared with models pretrained with the other two SSL approaches, the proposed SSL-Rank model consistently outperforms the rest of the models and is able to track the ground-truth degradation progression accurately under all scenarios with different ratios of labeled data, even when the labeled data is extremely sparse, as low as 1%. In general, since the labeled data are considered to be in the earlier degradation region, the estimations of the models fit better to the early-life ground-truth SOHs, while slight divergences can be observed toward the EOL. Nevertheless, the proposed SSL-Rank model is able to carry out an SOH estimation that not only is steady and smooth in early life, but also converges well to the ground-truth in late life, as shown in the zoomed-in sections. In comparison, the SSL-Recon model exhibits significantly higher oscillations in its estimation, and the estimation errors of the SSL-Recon and SSL-MO models become more pronounced in the deep-degradation regime. As for the ridge regression models, the Ridge-C model is able to build the overall trend of SOH from the input cycle number without any response to the occurring capacity recovery effect or fluctuations in the degradation process. Because of the lean structure, the Ridge-V model is able to carry out SOH estimation even with a limited amount of labeled data, in spite of significantly larger errors than the SSL models. However, as the labeled ratio increases, the Ridge-V model starts to indicate overfitting to the training data, as the estimation error in the lower SOH regime increases drastically.

Fig. 8 is the box plot of the absolute error of different models for SOH estimation under different ratios of labeled data. The aforementioned observations can be better confirmed in this visualization. The three SSL models achieve consistently good performance across all testing scenarios with varying proportions of labeled data. The SSL-Recon model and the SSL-MO model achieve comparable overall performance in most cases, though the latter typically produces more outliers. The proposed SSL-Rank model is able to outperform the rest of the models in every testing scenario, regardless of whether the labeled proportion is abundant or scarce, with fewer outliers in addition. The baseline SL model can still carry out relatively good SOH estimation when the amount of labeled data is sufficient, namely 50% or more, but is not able to learn the underlying degradation dynamics and thus fails to carry out SOH estimation when fewer labeled data are available. It is intriguing that even in cases where 100% or 50% labeled data is available, the SL model is still outperformed by all the SSL models, indicating that leveraging unlabeled charging data through SSL pretext tasks not only compensates for label scarcity, but also improves the model initialization and guides it toward a more favorable parameter space for downstream optimization, so that the model can make better use of the labeled data during fine-tuning and achieve superior SOH estimation performance even when labels are relatively abundant. As mentioned before, the Ridge-V model is able to survive insufficient labeled data because of its lean architecture, but it has consistently higher errors than all SSL models. In general, the Ridge-C model achieves comparable performance to the SSL-Recon and SSL-MO models, especially when the labeled ratio is smaller. However, it is still consistently outperformed by the proposed SSL-Rank model, indicating that models pretrained by the proposed ranking-based SSL approach are truly degradation-aware, instead of simply taking the cycle number as a proxy.

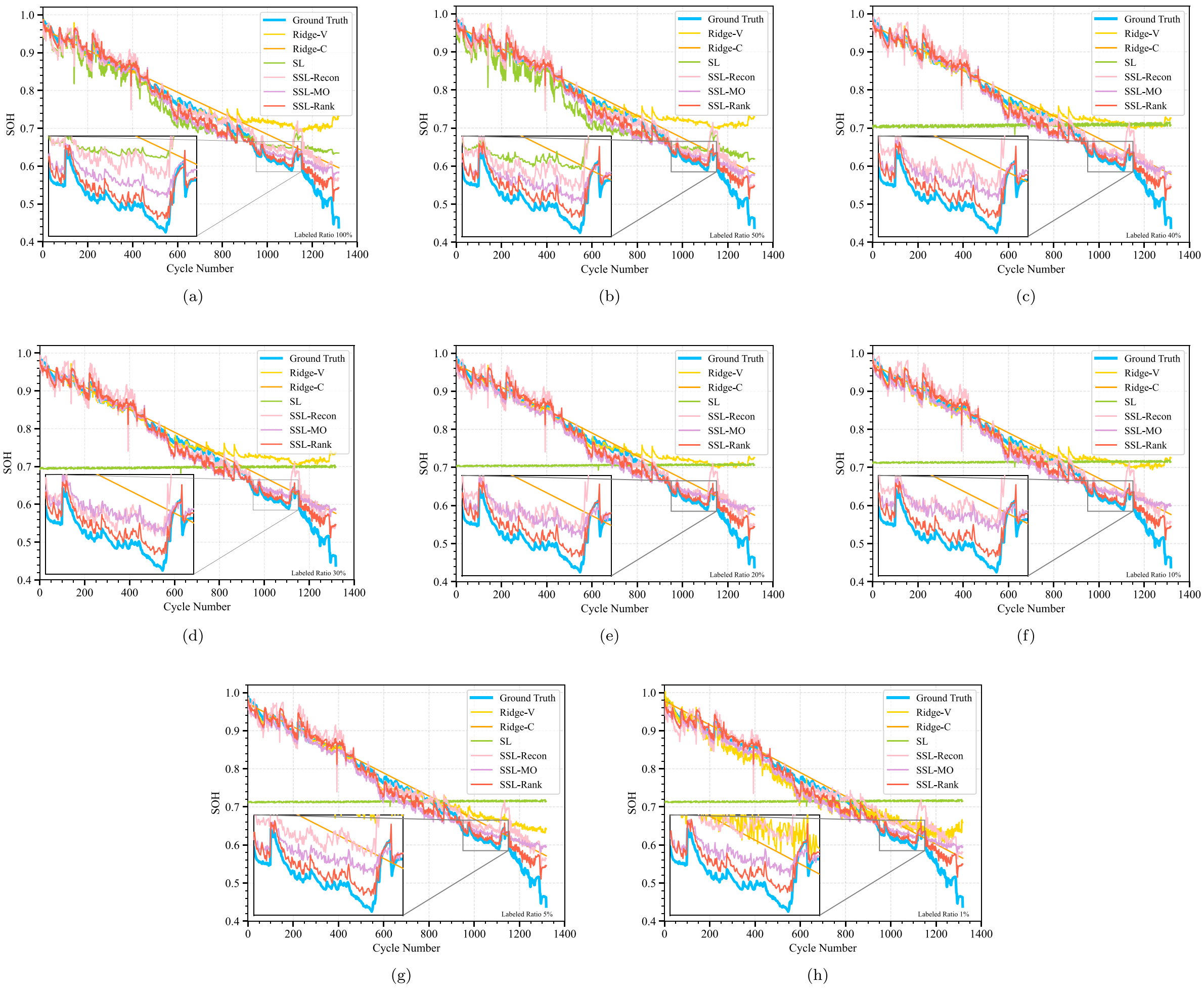

**Fig. 7.** SOH estimation results on the test cell CX2-34 of different models after fine-tuning or complete training on different ratios of labeled data. (a) 100% of labeled data available. (b) 50% of labeled data available. (c) 40% of labeled data available. (d) 30% of labeled data available. (e) 20% of labeled data available. (f) 10% of labeled data available. (g) 5% of labeled data available. (h) 1% of labeled data available.

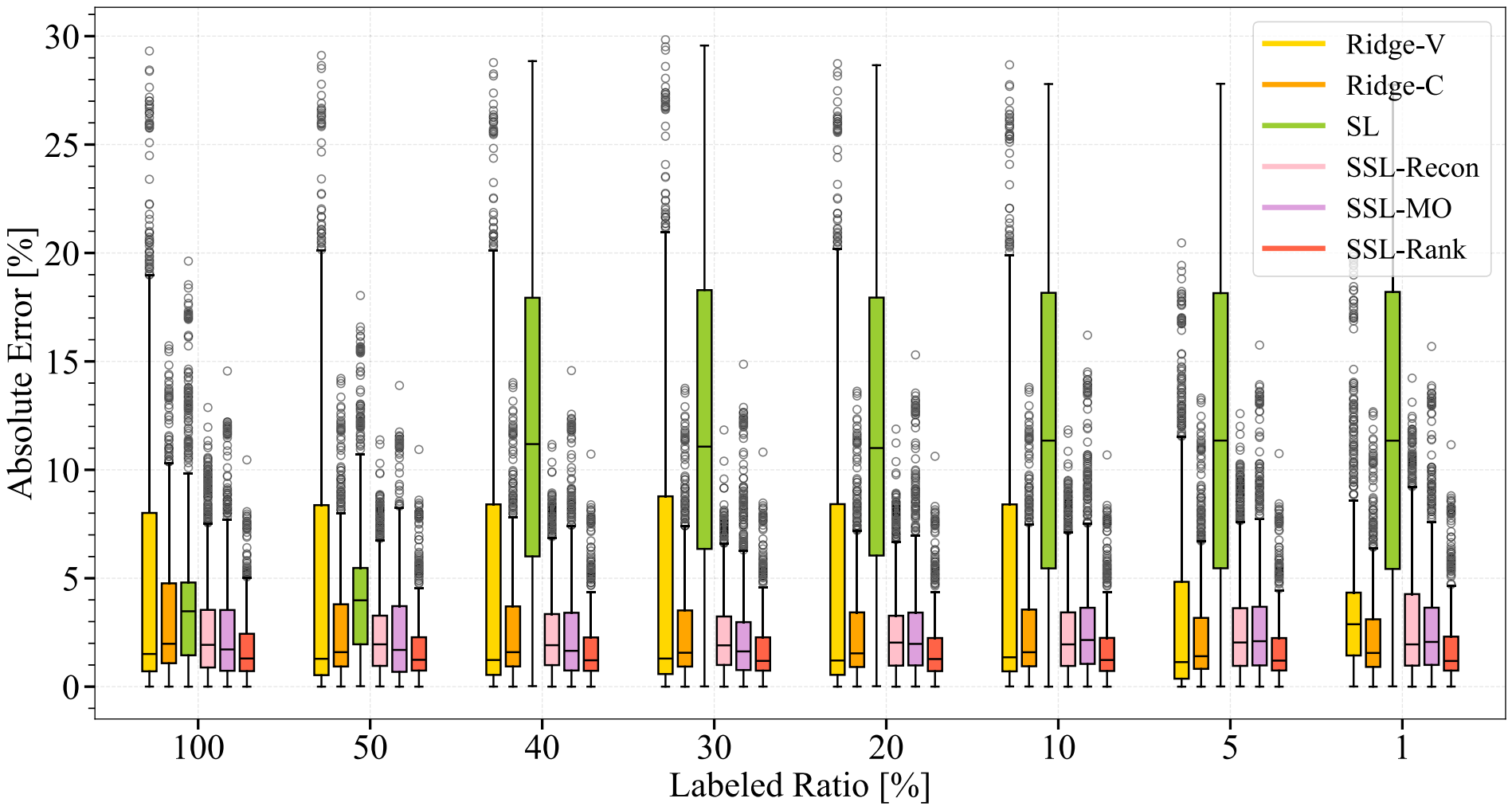

**Fig. 8.** Box plot of the absolute error for SOH estimation under different ratios of labeled data.

Table 3 summarizes the experiment results on different models' performance under varying ratios of labeled data above 80% SOH. The proposed SSL-Rank model leads across all four evaluation metrics in all testing scenarios with different labeled data ratios, achieving MAEs of 1.715%, 1.722%, 1.692%, and 1.718% for labeled data ratios of 100%, 50%, 10%, and 1%, respectively. The single-objective SSL-Recon model is outperformed by the dual-objective SSL-MO model in most test cases, except at 20% and 10% labeled data ratios, in terms of MAE. In the case when 30% of labeled data is available, the SSL-MO model achieves an MAE of 2.283%, which is lower than the MAE of 2.449% of the SSL-Recon model, while the RMSE of 3.311% of the SSL-MO model is higher than the RMSE of 3.169% of the SSL-Recon model, indicating that the estimation of the SSL-MO model contains more outliers. The SL model achieves MAEs of 4.017% and 4.031% in test scenarios with labeled data ratios of 100% and 50%, respectively, but collapses when fewer labeled data are available, with MAEs around 12%. In all test scenarios, the proposed SSL-Rank model is able to maintain an $R^2$ of over 0.970, indicating its strong performance in SOH regression. The MAXs of around 10% typically characterize the final estimation error in the deepest-degradation regime due to significantly accelerated aging in this stage, suggesting that late-life labeled data, even if sparse, remain important for reliable deployment in real-world applications. The error metrics of the two ridge regression models also confirm our aforementioned observations that models pretrained using the proposed SSL approach not only outperform intuitive, lean machine learning baselines, but also are truly degradation-aware, instead of simply taking the cycle number as a proxy.

**Table 3**
Comparison of different models' performance under varying ratios of labeled data above 80% SOH.

| Labeled ratio [%] | Model | Evaluation metrics | | | |
|---|---|---|---|---|---|
| | | MAE [%] | RMSE [%] | $R^2$ | MAX [%] |
| 100 | SSL-Rank | **1.715** | **2.255** | **0.972** | **10.458** |
| | SSL-Recon | 2.768 | 3.789 | 0.921 | 12.874 |
| | SSL-MO | 2.409 | 3.407 | 0.936 | 14.556 |
| | SL | 4.017 | 5.384 | 0.841 | 19.625 |
| | Ridge-C | 3.172 | 4.421 | 0.892 | 15.722 |
| | Ridge-V | 4.708 | 7.836 | 0.662 | 29.315 |
| 50 | SSL-Rank | **1.722** | **2.302** | **0.971** | **10.937** |
| | SSL-Recon | 2.557 | 3.366 | 0.938 | 11.380 |
| | SSL-MO | 2.370 | 3.303 | 0.940 | 13.889 |
| | SL | 4.301 | 5.381 | 0.841 | 18.038 |
| | Ridge-C | 2.627 | 3.738 | 0.923 | 14.213 |
| | Ridge-V | 4.780 | 8.038 | 0.645 | 29.112 |
| 40 | SSL-Rank | **1.699** | **2.259** | **0.972** | **10.724** |
| | SSL-Recon | 2.590 | 3.401 | 0.936 | 11.168 |
| | SSL-MO | 2.362 | 3.370 | 0.938 | 14.576 |
| | SL | 12.203 | 14.317 | −0.127 | 28.848 |
| | Ridge-C | 2.584 | 3.666 | 0.926 | 14.021 |
| | Ridge-V | 4.761 | 7.993 | 0.649 | 28.781 |
| 30 | SSL-Rank | **1.707** | **2.279** | **0.971** | **10.815** |
| | SSL-Recon | 2.449 | 3.169 | 0.945 | 11.838 |
| | SSL-MO | 2.283 | 3.311 | 0.940 | 14.869 |
| | SL | 12.442 | 14.593 | −0.171 | 29.566 |
| | Ridge-C | 2.510 | 3.560 | 0.930 | 13.759 |
| | Ridge-V | 4.996 | 8.358 | 0.616 | 29.837 |
| 20 | SSL-Rank | **1.692** | **2.241** | **0.972** | **10.624** |
| | SSL-Recon | 2.564 | 3.350 | 0.938 | 11.878 |
| | SSL-MO | 2.597 | 3.588 | 0.929 | 15.300 |
| | SL | 12.184 | 14.290 | −0.123 | 28.659 |
| | Ridge-C | 2.471 | 3.505 | 0.932 | 13.625 |
| | Ridge-V | 4.774 | 8.002 | 0.648 | 28.733 |
| 10 | SSL-Rank | **1.692** | **2.249** | **0.972** | **10.686** |
| | SSL-Recon | 2.655 | 3.528 | 0.932 | 11.839 |
| | SSL-MO | 2.803 | 3.873 | 0.918 | 16.210 |
| | SL | 11.966 | 14.039 | −0.084 | 27.791 |
| | Ridge-C | 2.533 | 3.584 | 0.929 | 13.802 |
| | Ridge-V | 4.797 | 7.910 | 0.656 | 28.679 |
| 5 | SSL-Rank | **1.693** | **2.258** | **0.972** | **10.749** |
| | SSL-Recon | 2.806 | 3.789 | 0.921 | 12.592 |
| | SSL-MO | 2.721 | 3.756 | 0.922 | 15.753 |
| | SL | 11.969 | 14.043 | −0.085 | 27.803 |
| | Ridge-C | 2.330 | 3.350 | 0.938 | 13.308 |
| | Ridge-V | 3.040 | 5.139 | 0.855 | 20.464 |
| 1 | SSL-Rank | **1.718** | **2.329** | **0.970** | **11.157** |
| | SSL-Recon | 3.075 | 4.293 | 0.899 | 14.228 |
| | SSL-MO | 2.689 | 3.707 | 0.924 | 15.689 |
| | SL | 11.957 | 14.029 | −0.082 | 27.749 |
| | Ridge-C | 2.307 | 3.187 | 0.944 | 12.681 |
| | Ridge-V | 3.634 | 5.054 | 0.860 | 22.066 |

The results of this experiment showcase the effectiveness and robustness of the proposed SSL framework across different labeled data proportions under the practical earlier-life labeling setting. Even with an extremely limited amount of labeled data, the proposed method is still able to maintain a high estimation accuracy, further demonstrating that the self-supervised pretraining stage has successfully endowed the model with informative and degradation-aligned prior knowledge that can be effectively transferred to downstream SOH estimation under label sparsity in both quantity and coverage.

### *4.3. Analysis on the influence of label distribution*

In this experiment, we aim to evaluate the SSL-Rank model's performance under different distributions of sparsely labeled data and thereby analyze the influence of label distribution on the task of SOH estimation. Since the test results in Section 4.2 already show that the model pretrained utilizing our proposed ranking-based SSL approach is able to carry out accurate and robust SOH estimation after fine-tuning, even only on an extremely small number of labeled data, we explore the different distributions on rather small labeled ratios here, namely 10%, 5%, 2%, and 1% of the whole training set, where 10% corresponds to approximately 260 samples. The four studied types of distribution are uniform, random, early-only, and late-only. For the uniform setting, the labeled samples are selected to cover the entire available SOH range as evenly as possible. For the random setting, the labeled samples are randomly drawn from all available data. In the early-only setting, only samples from the extreme early degradation stage, namely above 85% SOH, are retained as labeled data, whereas in the late-only setting, only samples from the late degradation stage, namely below 60% SOH, are labeled. This experiment is designed as a supplementary analysis to the main experiment in Section 4.2, with the focus shifted from the overall effectiveness of the proposed SSL framework to the specific influence of label distribution of the training set under very limited labeling budgets.

Fig. 9 shows the regression performance of the proposed SSL-Rank model fine-tuned with 10% or 5% of labeled data sampled from different distributions. At first glance, we can see that the type of distribution of the labeled data has some minor influence on the overall regression performance. The $R^2$ decreases from 0.979 to 0.977 when 10% of the labeled data are sampled randomly from the whole lifespan instead of uniformly, and decreases from 0.981 to 0.977 when the labeled ratio is 5%, indicating that a more even coverage of the degradation trajectory is still beneficial. The early-only distribution, on the other hand, seems to yield worse training outcomes: the $R^2$ is 0.974 when the labeled ratio is 10%, whereas the model fails the SOH estimation task when the ratio decreases to 5%. In fact, we have also tried to sample 10% of the labeled data only from the range above 90%, where the model fails as well. The deterioration in model performance when labeled data are only available for the extremely early life stage is due to the limited diversity of charging curves in this stage. As shown in Fig. 5, the pattern shift in the voltage curve during CC charging is minor in the early-degradation stage, but it accelerates significantly toward the late-degradation stage, meaning that labeled samples drawn from the early-only distribution are not diverse enough and thus the model can hardly generalize on such data. The comparison with the test scenario using 1% labeled data in Section 4.2 also confirms this assumption, as simply extending the SOH threshold from 85% to 80% significantly alleviates the generalization problem, even with much fewer samples. However, at the same time, we also observe that the compensation for the lack of data diversity by larger data quantity also exists to some degree, as showcased by the regression performance under 10% and 5% of early-only labeled data. On the contrary, the late-only distribution leads to the overall performance even better than the uniform distribution. A plausible reason is that the voltage curves in this stage exhibit much more pronounced aging-related variations and pattern shifts, so that the labeled samples provide stronger supervisory signals for calibrating the pretrained model to the SOH regression task. In addition, a major source of error for uniform and random distributions is the divergence between the estimation and the ground truth toward the EOL, where significantly accelerated aging happens. This is also compensated for when more labeled samples in the late-degradation region are available for model fine-tuning. The label-distribution analysis suggests that, if a few additional labels can be obtained, they should preferably improve coverage of poorly represented degradation regions instead of being concentrated only in the early-life regime.

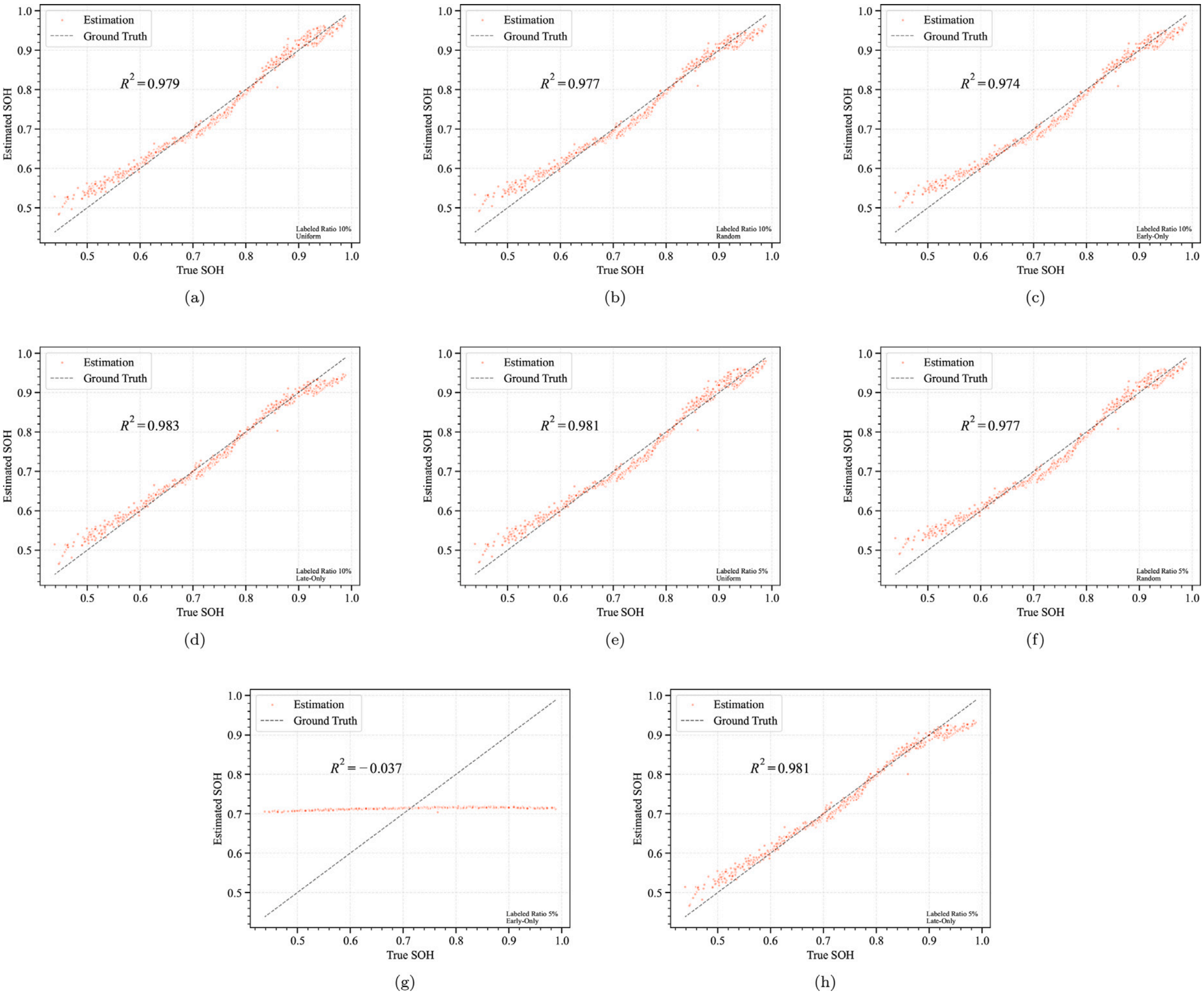

**Fig. 9.** Regression performance of the proposed SSL-Rank model fine-tuned with different ratios and distributions of labeled data. (a) 10% labeled data, uniform distribution. (b) 10% labeled data, random distribution. (c) 10% labeled data, early-only distribution. (d) 10% labeled data, late-only distribution. (e) 5% labeled data, uniform distribution. (f) 5% labeled data, random distribution. (g) 5% labeled data, early-only distribution. (h) 5% labeled data, late-only distribution.

Fig. 10 visualizes the MAEs of the SSL-Rank model's estimation with different ratios and distributions of labeled data for fine-tuning across different SOH windows. In general, the estimation error in the SOH range below 60% is significantly higher than in other SOH windows for each case due to the accelerated late-life degradation. The model fails the SOH estimation task when fine-tuned on less than 5% of labeled data under early-only distribution, causing the different scales of the bar graphs. In Fig. 10(a), we can clearly observe an increasing trend in the SOH estimation error in the late-life SOH window below 60% from uniform distribution to random distribution and then to early-only distribution. Sampling labeled data from the late-only distribution increases the estimation error in the early-life SOH window above 90%, but at the same time significantly decreases the estimation error in the late-life SOH window below 60% as well as in the middle-life SOH windows from 60% to 90%, resulting in an improvement in the overall performance.

Table 4 summarizes the estimation results of the proposed SSL-Rank model fine-tuned under different distributions of sparsely labeled data. Many of the aforementioned observations can be confirmed. The model fine-tuned with late-only labeled data outperforms the rest of the label distributions consistently in terms of MAE, with the MAEs of 1.320%, 1.369%, 1.369%, and 1.350% for cases where 10%, 5%, 2%, and 1% of the data are labeled, respectively. This suggests that, under the same labeling budget, label coverage over the more informative late-degradation region is particularly beneficial for downstream SOH estimation, which indicates that, when only a very limited number of labels can be acquired, prioritizing samples from the later degradation stage may be more advantageous than distributing labels evenly or concentrating them in the early-life stage. The failure of the model fine-tuned with early-only labeled data in cases of fewer labeled data shows that the lack of data diversity and data quantity jointly affect the effectiveness of downstream fine-tuning. However, to some extent, these two factors can compensate for each other: limited diversity may be partially alleviated by a larger number of labeled samples, while limited quantity may be mitigated if the labeled data cover more informative and diverse degradation stages. These findings further confirm the effectiveness of the proposed ranking-based SSL strategy, which enables the model to capture degradation-relevant information from unlabeled data and thus alleviates the dependence of downstream fine-tuning on both label quantity and label diversity.

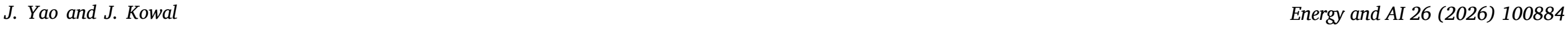

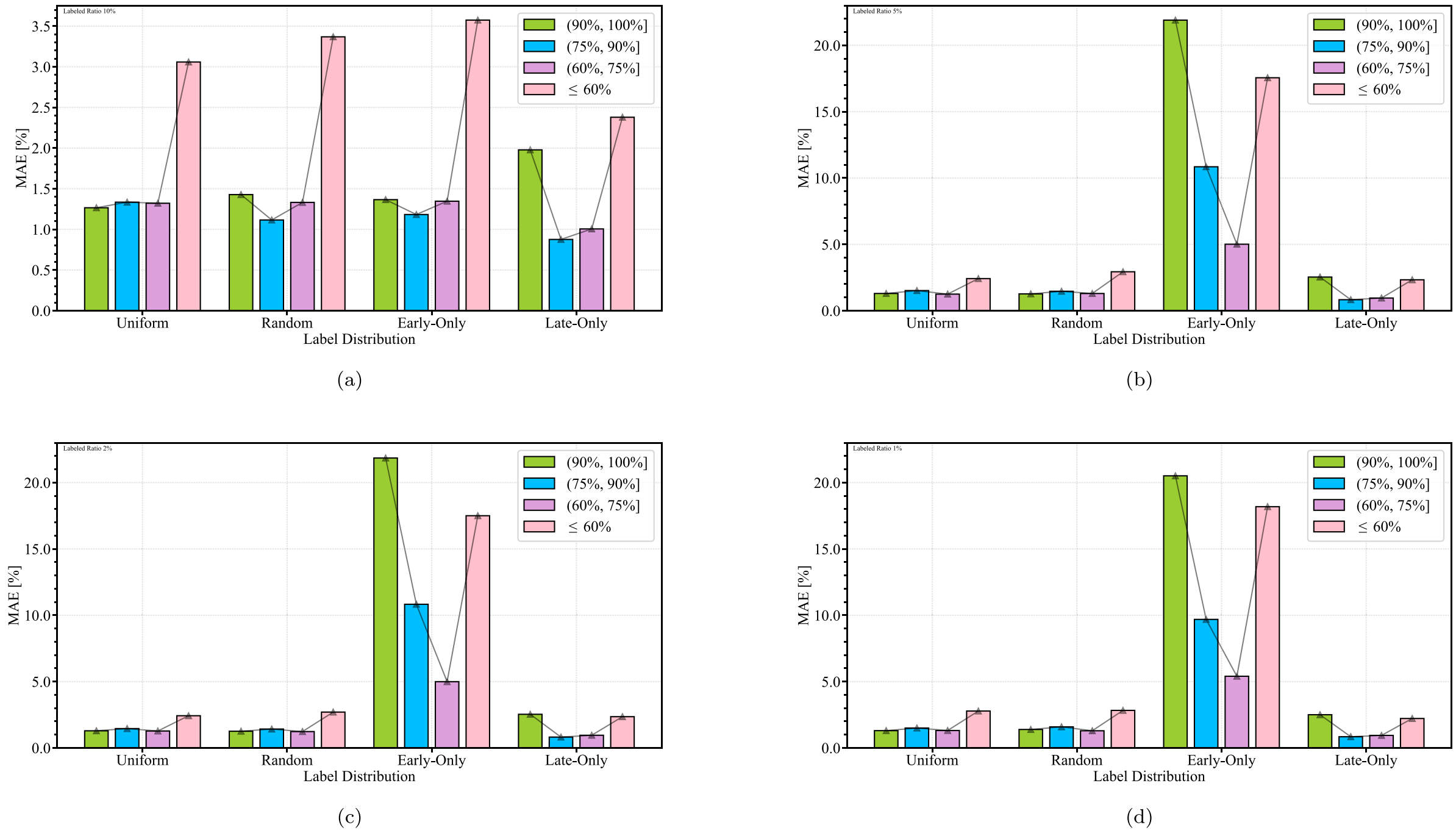

**Fig. 10.** Bar graph of the MAEs of the SSL-Rank model's estimation with different ratios and distributions of labeled data for fine-tuning across SOH windows. (a) 10% of labeled data available. (b) 5% of labeled data available. (c) 2% of labeled data available. (d) 1% of labeled data available.

**Table 4**
Evaluation of the SSL-Rank model's performance under different distributions of sparsely labeled data.

| Labeled ratio [%] | Distribution | Evaluation metrics | | | |
|---|---|---|---|---|---|
| | | MAE [%] | RMSE [%] | $R^2$ | MAX [%] |
| 10 | Uniform | 1.569 | 1.946 | 0.979 | 9.032 |
| | Random | 1.566 | 2.037 | 0.977 | 9.515 |
| | Early-Only | 1.615 | 2.160 | 0.974 | 10.058 |
| | Late-Only | **1.320** | **1.744** | **0.983** | **7.605** |
| 5 | Uniform | 1.519 | **1.858** | **0.981** | 7.724 |
| | Random | 1.586 | 2.032 | 0.977 | 9.199 |
| | Early-Only | 11.682 | 13.728 | −0.037 | 27.632 |
| | Late-Only | **1.369** | 1.862 | **0.981** | **7.565** |
| 2 | Uniform | 1.508 | **1.804** | **0.982** | **7.480** |
| | Random | 1.516 | 1.935 | 0.979 | 8.791 |
| | Early-Only | 11.654 | 13.694 | −0.031 | 27.572 |
| | Late-Only | **1.369** | 1.863 | 0.981 | 7.659 |
| 1 | Uniform | 1.589 | 1.904 | 0.980 | 8.054 |
| | Random | 1.635 | 2.013 | 0.978 | 8.753 |
| | Early-Only | 11.266 | 13.240 | 0.036 | 27.212 |
| | Late-Only | **1.350** | **1.839** | **0.981** | **7.420** |

### *4.4. Robustness analysis across data splits and fine-tuning strategies*

In this section, we train, validate, and test the models' performance across different data splits in order to further test the robustness of the proposed SSL framework under different test scenarios, where each cell is used as the test cell in turn, while the rest are used for training and validation. Specifically, we report results from four different data splits:

1. CX2-34 as test cell: CX2-37 and CX2-38 for training, CX2-36 for validation.
2. CX2-36 as test cell: CX2-37 and CX2-38 for training, CX2-34 for validation.
3. CX2-37 as test cell: CX2-34 and CX2-38 for training, CX2-36 for validation.
4. CX2-38 as test cell: CX2-34 and CX2-37 for training, CX2-36 for validation.

Based on the depth of degradation, the four different data splits represent different test scenarios: Since CX2-37 has the shallowest degradation, the SOH estimation task is relatively easy, considering its degradation process is completely covered by the provided training data; since CX2-38 has the deepest degradation, the SOH estimation task becomes a difficult extrapolation, as its degradation regime is not covered by the training cells; since the degradation of CX2-34 and CX2-36 is enclosed by the training distribution, the difficulty of their SOH estimation is rather moderate.

For each test scenario, the sparsely labeled situations are considered, with the ratio of labeled data ranging from 10% to 1%. In each case, three models are trained for comparison. Besides the strong machine learning baseline for such data-scarce scenarios, namely the Ridge-V model, we further fine-tune the SSL-Rank-pretrained model using two different fine-tuning strategies to study their influence: only fine-tune the task head with the pretrained encoder frozen (Head), and fine-tune the full model including the pretrained encoder and the task head (Full).

The results are shown in Table 5. Overall, the SSL-Rank model with the head fine-tuned is able to carry out the SOH estimation task on sparsely labeled data successfully in different test scenarios, where it achieves the lowest error in most cases. For the simpler test case of CX2-37, the Ridge-V model can achieve comparable performance to the SSL-Rank model with the head fine-tuned, with the MAE of 1.307% when 1% of labeled data is available and the MAE of 1.676% when 5% of labeled data is available. As mentioned before, this is due to the fact that the degradation process of the test cell is completely covered by the training samples, leading to a relatively naive test case that can be easily solved without the help of extra information or pretraining. On the contrary, when it comes to the extrapolation case where CX2-38 is used as the test cell, the SSL-Rank model with the head fine-tuned still outperforms the Ridge-V model consistently due to the extra degradation-aware knowledge gained through the ranking-based SSL pretraining, with moderately increased MAEs of 3.195% when 1% of labeled data is available and 3.885% when 2% of labeled data is available, even when this test scenario is extremely difficult due to the extrapolation. Although the MAX has increased drastically, we notice that it is caused by one estimation outlier in the deep-degradation regime, which could possibly be mitigated through more delicate pretraining. For the two moderate test cases where CX2-34 and CX2-36 are respectively used as the test cell, the SSL-Rank model with the head fine-tuned is able to carry out robust SOH estimation with different labeled ratios as well. As for the fully fine-tuned SSL-Rank model, it fails the SOH estimation task in all cases with very high errors. Not surprisingly, when the pretrained model is fully fine-tuned, the knowledge the encoder gained through the pretraining is undermined, as the full fine-tuning enforces the biased coverage of degradation information onto the encoder, which should have kept its awareness of the overall degradation process in order for an accurate SOH estimation over the entire lifespan.

**Table 5**
Evaluation of the SSL-Rank model's performance under different data splits using different fine-tuning strategies with sparsely labeled data.

| Test cell | Labeled ratio [%] | Strategy | Evaluation metrics | | | |
|---|---|---|---|---|---|---|
| | | | MAE [%] | RMSE [%] | $R^2$ | MAX [%] |
| CX2-34 | 10 | Ridge-V | 4.797 | 7.910 | 0.656 | 28.679 |
| | | Head | **1.692** | **2.249** | **0.972** | **10.686** |
| | | Full | 9.180 | 10.903 | 0.346 | 26.325 |
| | 5 | Ridge-V | 3.040 | 5.139 | 0.855 | 20.464 |
| | | Head | **1.693** | **2.258** | **0.972** | **10.749** |
| | | Full | 9.277 | 10.988 | 0.336 | 26.184 |
| | 2 | Ridge-V | 4.219 | 6.713 | 0.752 | 27.101 |
| | | Head | **1.689** | **2.260** | **0.972** | **10.860** |
| | | Full | 9.075 | 10.817 | 0.356 | 26.594 |
| | 1 | Ridge-V | 3.634 | 5.054 | 0.860 | 22.066 |
| | | Head | **1.718** | **2.329** | **0.970** | **11.157** |
| | | Full | 9.122 | 10.837 | 0.354 | 26.340 |
| CX2-36 | 10 | Ridge-V | 4.926 | 7.063 | 0.691 | 26.685 |
| | | Head | **2.430** | **3.207** | **0.936** | **13.590** |
| | | Full | 11.028 | 12.962 | −0.040 | 30.678 |
| | 5 | Ridge-V | 3.326 | 4.631 | 0.867 | 18.821 |
| | | Head | **2.401** | **3.183** | **0.937** | **13.565** |
| | | Full | 10.893 | 13.001 | −0.046 | 28.773 |
| | 2 | Ridge-V | 4.240 | 6.062 | 0.773 | 24.834 |
| | | Head | **2.400** | **3.177** | **0.938** | **13.529** |
| | | Full | 10.907 | 12.874 | −0.026 | 29.962 |
| | 1 | Ridge-V | 3.215 | 4.110 | 0.895 | 16.770 |
| | | Head | **2.459** | **3.232** | **0.935** | **13.589** |
| | | Full | 10.973 | 12.866 | −0.025 | 30.898 |
| CX2-37 | 10 | Ridge-V | 2.236 | 3.560 | 0.869 | 10.949 |
| | | Head | **1.811** | **2.216** | **0.949** | **6.569** |
| | | Full | 11.242 | 13.433 | −0.871 | 28.453 |
| | 5 | Ridge-V | **1.676** | 2.548 | 0.933 | 8.273 |
| | | Head | 1.767 | **2.136** | **0.953** | **5.974** |
| | | Full | 11.414 | 13.626 | −0.925 | 28.725 |
| | 2 | Ridge-V | 2.544 | 3.748 | 0.854 | 10.909 |
| | | Head | **1.892** | **2.291** | **0.946** | **6.092** |
| | | Full | 11.087 | 13.268 | −0.825 | 28.255 |
| | 1 | Ridge-V | **1.307** | **1.572** | **0.974** | **7.429** |
| | | Head | 2.116 | 2.578 | 0.931 | 7.706 |
| | | Full | 11.593 | 13.826 | −0.982 | 29.005 |
| CX2-38 | 10 | Ridge-V | 5.332 | 9.438 | 0.741 | 36.112 |
| | | Head | **3.994** | **5.761** | **0.904** | **31.126** |
| | | Full | 14.935 | 18.332 | 0.024 | 48.069 |
| | 5 | Ridge-V | 4.670 | 8.273 | 0.801 | 32.201 |
| | | Head | **3.855** | **5.625** | **0.908** | **30.978** |
| | | Full | 14.914 | 18.404 | 0.016 | 49.315 |
| | 2 | Ridge-V | 5.407 | 10.446 | 0.683 | 43.189 |
| | | Head | **3.885** | **5.796** | **0.902** | **31.266** |
| | | Full | 14.757 | 18.252 | 0.032 | 49.382 |
| | 1 | Ridge-V | 6.540 | 13.039 | 0.506 | 53.256 |
| | | Head | **3.195** | **5.084** | **0.925** | **30.000** |
| | | Full | 14.838 | 18.312 | 0.026 | 49.083 |

This experiment showcases the robustness of the proposed SSL framework across different test scenarios under label sparsity, no matter if it is a naive case where the target degradation is shallow, or it is a difficult extrapolation case where the target degradation is even much deeper than the training data.

### 4.5. Study on industrial applicability

In this section, a concise study on the industrial applicability of the proposed SSL framework for robust SOH estimation under label sparsity is conducted. As a matter of fact, the computationally intensive part of the proposed framework is the ranking-based SSL pretraining, which can be done offline or on the cloud without the limitations of the edge BMS device. Therefore, the bottleneck for computational resources lies in the computational cost of the inference of the CNN-GRU model. We provide the key indicators that are critical for industrial applications of the proposed framework, including the number of floating-point operations (FLOPs), number of parameters, model size, and the average inference latency on the CPU. The number of parameters can be used to quantify the memory cost of the model, while the FLOPs can be used to quantify the computational cost of the model. The calculation of the FLOPs and the average inference latency is based on a hypothetical input voltage time series of length 300, just like in the previous experiments.

**Table 6**
Summary of the applicability key indicators of the proposed CNN-GRU model.

| FLOPs | Number of parameters | Model size [KB] | Inference latency [ms] |
|---|---|---|---|
| $4.3763 \times 10^6$ | 97 473 | 384.55 | 15.323 |

As shown in Table 6, the proposed CNN-GRU model contains 97473 trainable parameters, corresponding to a model size of only 384.55 KB. For an input CC voltage sequence with a length of 300, one forward pass requires $4.3763 \times 10^6$ FLOPs. The measured average inference latency on a general-purpose CPU is 15.323 ms. This value should be interpreted as a reference measurement rather than a direct estimate of the latency on an automotive BMS microcontroller, where the available computational resources are typically more limited. Nevertheless, SOH estimation is usually a low-frequency diagnostic task and does not require the millisecond-level update rate needed by typical safety protection functions. Therefore, the relatively small parameter count and moderate FLOPs suggest that the deployed inference model is computationally lightweight, especially for resource-capable onboard computing units or cloud-assisted BMS platforms. Further benchmarking on target embedded BMS hardware is still required before practical deployment.

## 5. Conclusion

An accurate estimation of SOH underpins a safe and optimized use of the battery system. Although compelling, data-driven SOH estimation models rely on large amounts of high-quality labeled cycling data, where the SOH labels are typically obtained through standardized checkup tests that cover the entire lifespan of the batteries under controlled conditions. However, in practical application scenarios, battery aging data often face the problem of label sparsity in both quantity and coverage due to the time-consuming and costly nature of such lifespan-wide calibration tests, while large volumes of unlabeled operational cycling data are available. Therefore, in this work, we propose a degradation-aligned SSL framework that learns aging-consistent representations from unlabeled CC charging curves through a cycle-order ranking objective as the pretext task for pretraining, thereby enabling robust SOH estimation after fine-tuning on only a limited amount of labeled data for practical scenarios where data labels are sparse in both quantity and coverage. In addition, we develop a CNN-GRU model that extracts local patterns from charging curves via convolutional layers and then integrates these features sequentially through recurrent units for accurate SOH estimation as well as self-supervised pretraining. Comprehensive experiments are conducted on the CALCE battery aging dataset to evaluate the model performance after self-supervised pretraining and after fine-tuning on sparsely labeled data. Test results not only demonstrate that the proposed ranking-based SSL approach proves to endow the pretrained model with degradation awareness from unlabeled data, where Spearman's correlation coefficients between the learned aging scores and the ground-truth SOHs are extremely close to −1, but also showcase that the proposed SSL-Rank model can carry out accurate, robust SOH estimation after fine-tuning, even when only an extremely limited amount of 1% of unevenly distributed labeled training data is available, where an MAE of 1.718% and an RMSE of 2.329% can be achieved on the test cell. In addition, in-depth analyses are presented regarding the influences of label distribution of battery degradation data and cross-cell robustness. So far, only one cell model under one cycling condition in the laboratory-based scenario has been studied. As a result, the present evaluation mainly supports the effectiveness of the proposed framework in a controlled label-sparse setting, but does not fully establish its generalization ability to different cell chemistries, temperatures, or operating protocols. Therefore, it would be more than meaningful for future work to extend the proposed SSL framework to cross-cell and cross-condition settings that better reflect the variability encountered in real-world applications. In addition, a more systematic sensitivity analysis over $d_{min}$ would be beneficial for the implementation of the proposed framework. Still, the relatively large MAX in the deep-degradation region suggests that further validation and safety-oriented calibration are required before deployment. In particular, uncertainty-aware extensions could help identify predictions in poorly labeled or deeply degraded regions where the model may be less reliable, and could therefore make the framework more suitable for safety-critical BMS deployment. Considering the practical motivation and the encouraging experimental results, we believe this work could shed new light on label-efficient SOH estimation of lithium-ion batteries, addressing a practical need in battery management.

## Abbreviations

The following abbreviations are used in this manuscript:

| | |
|---|---|
| BMS | Battery Management System |
| BOL | Beginning of Life |
| CALCE | Center for Advanced Life Cycle Engineering |
| CC | Constant Current |
| CV | Constant Voltage |
| CNN | Convolutional Neural Network |
| DNN | Deep Neural Network |
| ECM | Equivalent Circuit Model |
| EKF | Extended Kalman Filter |
| EOL | End of Life |
| EV | Electric Vehicle |
| FC | Fully Connected |
| FLOPs | Floating-Point Operations |
| GRU | Gated Recurrent Unit |
| ICA | Incremental Capacity Analysis |
| LAM | Loss of Active Material |
| LCO | Lithium Cobalt Oxide |
| LLI | Loss of Lithium Inventory |
| LSTM | Long Short-Term Memory |
| MAD | Median Absolute Deviation |
| MAE | Mean Absolute Error |
| MAX | Maximum Absolute Error |
| MLP | Multi-Layer Perceptron |
| NASA | National Aeronautics and Space Administration |
| ReLU | Rectified Linear Unit |
| RMSE | Root Mean Square Error |
| RNN | Recurrent Neural Network |
| RUL | Remaining Useful Life |
| SL | Supervised Learning |
| SOC | State of Charge |
| SOH | State of Health |
| SSL | Self-Supervised Learning |
| SSL-MO | Multi-Objective SSL |
| SSL-Rank | Ranking-Based SSL |
| SSL-Recon | Reconstruction-Based SSL |
| TPE | Tree-Structured Parzen Estimator |
| UKF | Unscented Kalman Filter |

## CRediT authorship contribution statement

**Jiaqi Yao:** Writing – review & editing, Writing – original draft, Visualization, Validation, Software, Resources, Methodology, Investigation, Formal analysis, Data curation, Conceptualization. **Julia Kowal:** Writing – review & editing, Supervision.

## Declaration of competing interest

The authors declare that they have no known competing financial interests or personal relationships that could have appeared to influence the work reported in this paper.

## Acknowledgment

We thank the High-Performance Computing Cluster of TU Berlin ZECM for the GPU resources and the Open Access Publication Fund of TU Berlin for the support.

## Data availability

The public dataset utilized in this work can be accessed through Ref. [59].

## References

[1] Hassan Q, Viktor P, J. Al-Musawi T, Mahmood Ali B, Algburi S, Alzoubi HM, Khudhair Al-Jiboory A, Zuhair Sameen A, Salman HM, Jaszczur M. The renewable energy role in the global energy transformations. Renew Energy Focus 2024;48:100545. http://dx.doi.org/10.1016/j.ref.2024.100545.

[2] Lin B, Li Z. Towards world's low carbon development: The role of clean energy. Appl Energy 2022;307:118160. http://dx.doi.org/10.1016/j.apenergy.2021.118160.

[3] Olabi A, Abdelkareem MA. Renewable energy and climate change. Renew Sustain Energy Rev 2022;158:112111. http://dx.doi.org/10.1016/j.rser.2022.112111.

[4] Feng Z, Yan B, Shen X, Zhang F, Tariq Z, Ouyang W, Han Z. A hybrid CNN-transformer surrogate model for the multi-objective robust optimization of geological carbon sequestration. Adv Water Resour 2025;196:104897. http://dx.doi.org/10.1016/j.advwatres.2025.104897.

[5] Roga S, Bardhan S, Kumar Y, Dubey SK. Recent technology and challenges of wind energy generation: A review. Sustain Energy Technol Assess 2022;52:102239. http://dx.doi.org/10.1016/j.seta.2022.102239.

[6] Pourasl HH, Barenji RV, Khojastehnezhad VM. Solar energy status in the world: A comprehensive review. Energy Rep 2023;10:3474–93. http://dx.doi.org/10.1016/j.egyr.2023.10.022.

[7] Xie J, Li Y-Z. Telemetry-driven physics-informed prediction for on-orbit thermal–electrical performance of solar panels in satellite power system. Aerosp Sci Technol 2026;177:112268. http://dx.doi.org/10.1016/j.ast.2026.112268.

[8] Yao J, Droese D, Kowal J. Joint online estimation of state of charge and internal temperature of lithium-ion batteries with multi-task learning. J Energy Storage 2026;155:121468. http://dx.doi.org/10.1016/j.est.2026.121468.

[9] Niu H, Zhang N, Lu Y, Zhang Z, Li M, Liu J, Zhang N, Song W, Zhao Y, Miao Z. Strategies toward the development of high-energy-density lithium batteries. J Energy Storage 2024;88:111666. http://dx.doi.org/10.1016/j.est.2024.111666.

[10] Horiba T, Maeshima T, Matsumura T, Koseki M, Arai J, Muranaka Y. Applications of high power density lithium ion batteries. J Power Sources 2005;146(1–2):107–10. http://dx.doi.org/10.1016/j.jpowsour.2005.03.205.

[11] Nasajpour-Esfahani N, Garmestani H, Bagheritabar M, Jasim DJ, Toghraie D, Dadkhah S, Firoozeh H. Comprehensive review of lithium-ion battery materials and development challenges. Renew Sustain Energy Rev 2024;203:114783. http://dx.doi.org/10.1016/j.rser.2024.114783.

[12] Zhang J, Huang H, Zhang G, Dai Z, Wen Y, Jiang L. Cycle life studies of lithium-ion power batteries for electric vehicles: A review. J Energy Storage 2024;93:112231. http://dx.doi.org/10.1016/j.est.2024.112231.

[13] Liang Y, Zhao C-Z, Yuan H, Chen Y, Zhang W, Huang J-Q, Yu D, Liu Y, Titirici M-M, Chueh Y-L, Yu H, Zhang Q. A review of rechargeable batteries for portable electronic devices. InfoMat 2019;1(1):6–32. http://dx.doi.org/10.1002/inf2.12000.

[14] Zubi G, Dufo-López R, Carvalho M, Pasaoglu G. The lithium-ion battery: State of the art and future perspectives. Renew Sustain Energy Rev 2018;89:292–308. http://dx.doi.org/10.1016/j.rser.2018.03.002.

[15] Camargos PH, Dos Santos PHJ, Dos Santos IR, Ribeiro GS, Caetano RE. Perspectives on Li-ion battery categories for electric vehicle applications: A review of state of the art. Int J Energy Res 2022;46(13):19258–68. http://dx.doi.org/10.1002/er.7993.

[16] S. Rangarajan S, Sunddararaj SP, Sudhakar A, Shiva CK, Subramaniam U, Collins ER, Senjyu T. Lithium-ion batteries—The crux of electric vehicles with opportunities and challenges. Clean Technol 2022;4(4):908–30. http://dx.doi.org/10.3390/cleantechnol4040056.

[17] Diao W, Jiang J, Zhang C, Liang H, Pecht M. Energy state of health estimation for battery packs based on the degradation and inconsistency. Energy Procedia 2017;142:3578–83. http://dx.doi.org/10.1016/j.egypro.2017.12.248.

[18] Yao J, Zhao H, Kowal J. Fast-adaptive early-stage remaining useful life prediction of lithium-ion batteries with meta-learning. J Power Sources 2025;660:238569. http://dx.doi.org/10.1016/j.jpowsour.2025.238569.

[19] Jin S, Sui X, Huang X, Wang S, Teodorescu R, Stroe D-I. Overview of machine learning methods for lithium-ion battery remaining useful lifetime prediction. Electronics 2021;10(24):3126. http://dx.doi.org/10.3390/electronics10243126.

[20] Demirci O, Taskin S, Schaltz E, Acar Demirci B. Review of battery state estimation methods for electric vehicles-Part II: SOH estimation. J Energy Storage 2024;96:112703. http://dx.doi.org/10.1016/j.est.2024.112703.

[21] Xiong R, Li L, Tian J. Towards a smarter battery management system: A critical review on battery state of health monitoring methods. J Power Sources 2018;405:18–29. http://dx.doi.org/10.1016/j.jpowsour.2018.10.019.

[22] Rezvanizaniani SM, Liu Z, Chen Y, Lee J. Review and recent advances in battery health monitoring and prognostics technologies for electric vehicle (EV) safety and mobility. J Power Sources 2014;256:110–24. http://dx.doi.org/10.1016/j.jpowsour.2014.01.085.

[23] Chen L, Lü Z, Lin W, Li J, Pan H. A new state-of-health estimation method for lithium-ion batteries through the intrinsic relationship between ohmic internal resistance and capacity. Measurement 2018;116:586–95. http://dx.doi.org/10.1016/j.measurement.2017.11.016.

[24] Hu X, Jiang J, Cao D, Egardt B. Battery health prognosis for electric vehicles using sample entropy and sparse Bayesian predictive modeling. IEEE Trans Ind Electron 2015. http://dx.doi.org/10.1109/TIE.2015.2461523, 1–1.

[25] Ng KS, Moo C-S, Chen Y-P, Hsieh Y-C. Enhanced Coulomb counting method for estimating state-of-charge and state-of-health of lithium-ion batteries. Appl Energy 2009;86(9):1506–11. http://dx.doi.org/10.1016/j.apenergy.2008.11.021.

[26] Li X, Jiang J, Wang LY, Chen D, Zhang Y, Zhang C. A capacity model based on charging process for state of health estimation of lithium ion batteries. Appl Energy 2016;177:537–43. http://dx.doi.org/10.1016/j.apenergy.2016.05.109.

[27] Wang J, Liu P, Hicks-Garner J, Sherman E, Soukiazian S, Verbrugge M, Tataria H, Musser J, Finamore P. Cycle-life model for graphite-LiFePO4 cells. J Power Sources 2011;196(8):3942–8. http://dx.doi.org/10.1016/j.jpowsour.2010.11.134.

[28] Safari M, Delacourt C. Simulation-based analysis of aging phenomena in a commercial graphite/LiFePO4 cell. J Electrochem Soc 2011;158(12):A1436. http://dx.doi.org/10.1149/2.103112jes.

[29] Christensen J, Newman J. A mathematical model for the lithium-ion negative electrode solid electrolyte interphase. J Electrochem Soc 2004;151(11):A1977. http://dx.doi.org/10.1149/1.1804812.

[30] Kim I-S. A technique for estimating the state of health of lithium batteries through a dual-sliding-mode observer. IEEE Trans Power Electron 2010;25(4):1013–22. http://dx.doi.org/10.1109/TPEL.2009.2034966.

[31] Hosseininasab S, Lin C, Pischinger S, Stapelbroek M, Vagnoni G. State-of-health estimation of lithium-ion batteries for electrified vehicles using a reduced-order electrochemical model. J Energy Storage 2022;52:104684. http://dx.doi.org/10.1016/j.est.2022.104684.

[32] Plett GL. Extended Kalman filtering for battery management systems of LiPB-based HEV battery packs. J Power Sources 2004;134(2):277–92. http://dx.doi.org/10.1016/j.jpowsour.2004.02.033.

[33] Liu S, Dong X, Yu X, Ren X, Zhang J, Zhu R. A method for state of charge and state of health estimation of lithium-ion battery based on adaptive unscented Kalman filter. Energy Rep 2022;8:426–36. http://dx.doi.org/10.1016/j.egyr.2022.09.093.

[34] Schwunk S, Armbruster N, Straub S, Kehl J, Vetter M. Particle filter for state of charge and state of health estimation for lithium–Iron phosphate batteries. J Power Sources 2013;239:705–10. http://dx.doi.org/10.1016/j.jpowsour.2012.10.058.

[35] Ren Z, Du C. A review of machine learning state-of-charge and state-of-health estimation algorithms for lithium-ion batteries. Energy Rep 2023;9:2993–3021. http://dx.doi.org/10.1016/j.egyr.2023.01.108.

[36] Xia B, Ye M, Wei M, Wang Q, Lian G, Li Y. SOH estimation of lithium-ion batteries with local health indicators in multi-stage fast charging protocols. Energy 2025;334:137617. http://dx.doi.org/10.1016/j.energy.2025.137617.

[37] Chen Z, Peng Y, Shen J, Zhang Q, Liu Y, Zhang Y, Xia X, Liu Y. State of health estimation for lithium-ion batteries based on fragmented charging data and improved gated recurrent unit neural network. J Energy Storage 2025;115:115952. http://dx.doi.org/10.1016/j.est.2025.115952.

[38] Xia Z, Qahouq JAA. Lithium-ion battery ageing behavior pattern characterization and state-of-health estimation using data-driven method. IEEE Access 2021;9:98287–304. http://dx.doi.org/10.1109/ACCESS.2021.3092743.

[39] Zhang S, Zhai B, Guo X, Wang K, Peng N, Zhang X. Synchronous estimation of state of health and remaining useful lifetime for lithium-ion battery using the incremental capacity and artificial neural networks. J Energy Storage 2019;26:100951. http://dx.doi.org/10.1016/j.est.2019.100951.

[40] You G-W, Park S, Oh D. Diagnosis of electric vehicle batteries using recurrent neural networks. IEEE Trans Ind Electron 2017;64(6):4885–93. http://dx.doi.org/10.1109/TIE.2017.2674593.

[41] Fan Y, Xiao F, Li C, Yang G, Tang X. A novel deep learning framework for state of health estimation of lithium-ion battery. J Energy Storage 2020;32:101741. http://dx.doi.org/10.1016/j.est.2020.101741.

[42] Jaiswal A, Babu AR, Zadeh MZ, Banerjee D, Makedon F. A survey on contrastive self-supervised learning. Technologies 2020;9(1):2. http://dx.doi.org/10.3390/technologies9010002.

[43] Liu X, Zhang F, Hou Z, Mian L, Wang Z, Zhang J, Tang J. Self-supervised learning: Generative or contrastive. IEEE Trans Knowl Data Eng 2021. http://dx.doi.org/10.1109/TKDE.2021.3090866, 1–1.

[44] Gui J, Chen T, Zhang J, Cao Q, Sun Z, Luo H, Tao D. A survey on self-supervised learning: Algorithms, applications, and future trends. IEEE Trans Pattern Anal Mach Intell 2024;46(12):9052–71. http://dx.doi.org/10.1109/TPAMI.2024.3415112.

[45] Devlin J, Chang M-W, Lee K, Toutanova K. BERT: Pre-training of deep bidirectional transformers for language understanding. 2019, http://dx.doi.org/10.48550/arXiv.1810.04805, arXiv:1810.04805.

[46] Brown TB, Mann B, Ryder N, Subbiah M, Kaplan J, Dhariwal P, Neelakantan A, Shyam P, Sastry G, Askell A, Agarwal S, Herbert-Voss A, Krueger G, Henighan T, Child R, Ramesh A, Ziegler DM, Wu J, Winter C, Hesse C, Chen M, Sigler E, Litwin M, Gray S, Chess B, Clark J, Berner C, McCandlish S, Radford A, Sutskever I, Amodei D. Language models are few-shot learners. 2020, http://dx.doi.org/10.48550/arXiv.2005.14165, arXiv:2005.14165.

[47] Wu Z, Xiong Y, Yu SX, Lin D. Unsupervised feature learning via non-parametric instance discrimination. In: 2018 IEEE/CVF conference on computer vision and pattern recognition. Salt Lake City, UT: IEEE; 2018, p. 3733–42. http://dx.doi.org/10.1109/CVPR.2018.00393.

[48] Chen T, Kornblith S, Norouzi M, Hinton G. A simple framework for contrastive learning of visual representations. 2020, http://dx.doi.org/10.48550/arXiv.2002.05709, arXiv:2002.05709.

[49] Chen H, Wang Y, Lagadec B, Dantcheva A, Bremond F. Joint generative and contrastive learning for unsupervised person re-identification. In: 2021 IEEE/CVF conference on computer vision and pattern recognition (CVPR). Nashville, TN, USA: IEEE; 2021, p. 2004–13. http://dx.doi.org/10.1109/CVPR46437.2021.00204.

[50] Qi Z, Dong R, Fan G, Ge Z, Zhang X, Ma K, Yi L. Contrast with reconstruct: Contrastive 3D representation learning guided by generative pretraining. 2023, http://dx.doi.org/10.48550/arXiv.2302.02318, arXiv:2302.02318.

[51] Gidaris S, Singh P, Komodakis N. Unsupervised representation learning by predicting image rotations. 2018, http://dx.doi.org/10.48550/arXiv.1803.07728, arXiv:1803.07728.

[52] Agrawal P, Carreira J, Malik J. Learning to see by moving. 2015, http://dx.doi.org/10.48550/arXiv.1505.01596, arXiv:1505.01596.

[53] Che Y, Zheng Y, Sui X, Teodorescu R. Boosting battery state of health estimation based on self-supervised learning. J Energy Chem 2023;84:335–46. http://dx.doi.org/10.1016/j.jechem.2023.05.034.

[54] Wang T, Ma Z, Zou S, Chen Z, Wang P. Lithium-ion battery state-of-health estimation: A self-supervised framework incorporating weak labels. Appl Energy 2024;355:122332. http://dx.doi.org/10.1016/j.apenergy.2023.122332.

[55] Liang Q, Zhang M, Jin Y, Xia L, Liu C, Zhang T. Multi-level contrastive self-supervised learning with dynamic spectral-temporal embedding for state-of-health estimation of lithium-ion battery with limited labeled data. IEEE Trans Instrum Meas 2025. http://dx.doi.org/10.1109/TIM.2025.3614866, 1–1.

[56] Lecun Y, Bottou L, Bengio Y, Haffner P. Gradient-based learning applied to document recognition. Proc IEEE 1998;86(11):2278–324. http://dx.doi.org/10.1109/5.726791.

[57] Cho K, van Merrienboer B, Gulcehre C, Bahdanau D, Bougares F, Schwenk H, Bengio Y. Learning phrase representations using RNN encoder-decoder for statistical machine translation. 2014, http://dx.doi.org/10.48550/ARXIV.1406.1078.

[58] Hochreiter S, Schmidhuber J. Long short-term memory. Neural Comput 1997;9(8):1735–80. http://dx.doi.org/10.1162/neco.1997.9.8.1735.

[59] Battery Data |Center for Advanced Life Cycle Engineering. https://calce.umd.edu/battery-data.

[60] Zhang H, Gui X, Zheng S, Lu Z, Li Y, Bian J. BatteryML: An open-source platform for machine learning on battery degradation. 2023, http://dx.doi.org/10.48550/ARXIV.2310.14714.

[61] Akiba T, Sano S, Yanase T, Ohta T, Koyama M. Optuna: A next-generation hyperparameter optimization framework. 2019, http://dx.doi.org/10.48550/arXiv.1907.10902, arXiv:1907.10902.